\documentclass[prd,nofootinbib,showpacs,superscriptaddress, preprint]{revtex4-1}
\usepackage[T1]{fontenc}
\usepackage{amsmath,amssymb}
\usepackage{epsfig}
\usepackage{graphicx}
\usepackage{comment}
\usepackage{tikz}       
\usepackage{tikz-feynman}
\tikzfeynmanset{compat=1.1.0}

\begin{document}
\title{TeV Scale Leptogenesis via Dark Sector Scatterings}

\author{Debasish Borah}%
 \email{dborah@iitg.ac.in}
\affiliation{%
 Department of Physics, Indian Institute of Technology, Guwahati, Assam 781039, India
 }%
\author{Arnab Dasgupta}%
 \email{arnabdasgupta@protonmail.ch}
\affiliation{%
 School of Liberal Arts, Seoul-Tech, Seoul 139-743, Korea}%
\author{Sin Kyu Kang}%
 \email{skkang@seoultech.ac.kr}
\affiliation{%
 School of Liberal Arts, Seoul-Tech, Seoul 139-743, Korea}%
 
 \begin{abstract}
We propose a novel scenario of generating lepton asymmetry via annihilation and coannihilation of dark sector particles including  t-channel processes. In order to realistically implement this idea, we consider the scotogenic model having three right handed neutrinos and a new scalar doublet, all of which are odd under an in-built $Z_2$ symmetry. The lightest $Z_2$ odd particle, if electromagnetically neutral, can be a dark matter candidate while annihilation and coannihilation between different $Z_2$ odd particles into standard model leptons serve as the source of lepton asymmetry. The light neutrino masses arise at one loop level with $Z_2$ odd fields going inside the loop. We show that experimental data related to light neutrinos, dark matter relic abundance and baryon asymmetry can be simultaneously satisfied in the model for two different cases: one with fermion dark matter and the other with scalar dark matter. In both scenarios, t-channel annihilation as well as coannihilation of $Z_2$ odd particles play a non-trivial role in producing the non-zero CP asymmetry. Both the scenarios remain allowed from DM direct detection while keeping the scale of leptogenesis as low as TeV or less, lower than the one for vanilla leptogenesis scenario in scotogenic model along with the additional advantage of explaining the baryon-dark matter coincidence to some extent. Due to such low scale, the model is testable through rare decay experiments looking for charged lepton flavour violation.

\end{abstract}
\maketitle

\section{\label{sec:level1}Introduction}
There have been significant progress in last few decades in gathering evidences suggesting the presence of a mysterious, non-luminous form of matter, known as dark matter (DM) in the present universe, whose amount 
 is approximately five times more than the ordinary luminous or baryonic matter density $\Omega_B \approx 5\%$  \cite{Aghanim:2018eyx}.
%
Among different beyond standard model (BSM) proposals for DM, the weakly interacting massive particle (WIMP) paradigm remains the most widely studied scenario where a DM candidate typically with electroweak (EW) scale mass and interaction rate similar to EW interactions can give rise to the correct DM relic abundance, a remarkable coincidence often referred to as the \textit{WIMP Miracle}. 
On the other hand, out of equilibrium decay of a heavy particle leading to the generation of baryon asymmetry has been a very well known mechanism for baryogenesis \cite{Weinberg:1979bt, Kolb:1979qa}. One interesting way to implement such a mechanism is leptogenesis \cite{Fukugita:1986hr} where
a net leptonic asymmetry is generated first which gets converted into baryon asymmetry through $B+L$ violating EW sphaleron transitions.
The interesting feature of this scenario is that
the required lepton asymmetry can be generated 
within the framework of the seesaw mechanism 
that explains the origin of tiny neutrino masses \cite{Patrignani:2016xqp}, another observed phenomena which the SM fails to address.

Although these popular scenarios can explain the phenomena of DM and baryon asymmetry independently, it is nevertheless an interesting observation that DM and baryon abundance are very close to each other, within the same order of magnitudes $\Omega_{\rm DM} \approx 5 \Omega_{B}$. Discarding the possibility of any numerical coincidence, one is left with the task of constructing theories that can relate the origin of these two observed phenomena in a unified manner. There have been several proposals already which mainly fall into two broad categories. In the first one, the usual mechanism for baryogenesis  is extended to 
apply to the dark sector which is also asymmetric \cite{Nussinov:1985xr, Davoudiasl:2012uw, Petraki:2013wwa, Zurek:2013wia}. 
The second one is to produce such asymmetries through annihilations \cite{Yoshimura:1978ex, Yoshimura:1978ex, Barr:1979wb, Baldes:2014gca} where one or more particles involved in the annihilations eventually go out of thermal equilibrium in order to generate a net asymmetry. 
The so-called WIMPy baryogenesis \cite{Cui:2011ab, Bernal:2012gv, Bernal:2013bga} belongs to this category, where a dark matter particle freezes out to generate its own relic abundance and then
an asymmetry in the baryon sector is produced from DM annihilations. 
%
%

While there is no evidence yet for seesaw mechanism, recently the so-called scotogenic model \cite{Ma:2006km} as an alternative to canonical seesaw mechanism has been extensively studied, where
 Majorana light neutrino masses can be generated at one loop level with DM particle in the loop.
In the scotogenic model, 
 the required lepton asymmetry can be generated through right handed neutrino decays at a low scale $M_N \sim 10$ TeV at the cost of a strongly hierarchical neutrino Yukawa structure \cite{Racker:2013lua,Hugle:2018qbw}, but it can not explain the coincidence of baryon asymmetry and DM abundance. 

An interesting question raised is whether leptogenesis through (co-)annihilations of $Z_2$ odd particles can be realised by lowering the scale of leptogenesis further compared to vanilla leptogenesis in the scotogenic model. Giving an answer to this question is the main purpose of this work.
We examine how the (co-)annihilations of $Z_2$ odd particles can produce the lepton asymmetry while keeping the correct DM abundance,
and show that the DM relic abundance is correlated with baryon asymmetry in the scenario. 
To include all possible annihilations producing lepton asymmetry, we consider the annihilations and coannihilations of all $Z_2$ odd particles instead of restricting them to the lightest $Z_2$ odd particle which is also the DM candidate. If we consider the neutral component of the $Z_2$ odd scalar doublet as DM, then there exists s-channel coannihilation diagrams between DM and right handed neutrinos which can produce a net leptonic asymmetry. 
While the model satisfies correct DM abundance and lepton asymmetry, the DM sector can be probed at direct detection experiments as well as colliders due to the electroweak gauge interactions of scalar doublet DM. We then consider the fermion DM scenario where the lightest right handed neutrino plays the role of DM. While the annihilation of a pair of the fermionic DM can not produce a net lepton asymmetry in this case, the annihilation of $Z_2$ odd scalar doublets can contribute to the generation of lepton asymmetry.  Due to the natural absence of typical s-channel diagrams of scalar doublet annihilations leading to lepton asymmetry, here we show how t-channel diagrams (both tree level and one loop level) can play a non-trivial role in creating the required asymmetry. 

As far as we know, the contributions of (co-)annihilations of dark sector particles to lepton asymmetry in this minimal model was not considered before. In both the scenarios we address here, the criteria for "on-shell" -ness of loop particles in one loop annihilation diagrams dictate the particle spectrum and hence the nature of dark matter candidate. We show that it is possible to satisfy the requirement of baryon asymmetry, light neutrino mass and DM related constraints in both the scenarios while keeping the scale of leptogenesis as low as 5 TeV, lower than the scale of vanilla leptogenesis in the same model \cite{Hugle:2018qbw, Borah:2018rca}. Due to such a low scale, the model has another advantage to predict observable rates of charged lepton flavour violation accessible by the sensitivity of the future experiments.

This paper is arranged as follows. In section \ref{sec:level2}, we briefly review on the minimal scotogenic model followed by detailed discussion on leptogenesis from annihilation and coannihilations in this model in section \ref{sec:level3}. We finally conclude in section \ref{sec:level5}.

\section{\label{sec:level2}Minimal Scotogenic Model}
The minimal scotogenic model \cite{Ma:2006km} is the extension of the SM by three copies of right handed singlet neutrinos $N_i, i \in 1,2,3$ and one scalar field $\eta$ transforming as a doublet under $SU(2)_L$. An additional discrete symmetry $Z_2$ is incorporated under which these new fields are odd giving rise to the possibility of the lightest $Z_2$-odd particle being a suitable DM candidate. The Lagrangian involving the newly added singlet fermions is
\begin{equation}
{\cal L} \supset  \frac{1}{2}(M_N)_{ij} N_iN_j + \left(y_{ij} \, \bar{L}_i \tilde{\eta} N_j  + \text{h.c.} \right) \ . 
\end{equation}
The electroweak symmetry breaking occurs due to the non-zero vacuum expectation value (VEV) acquired by the neutral component of the SM Higgs doublet while the $Z_2$-odd doublet $\eta$ does not acquire any VEV.
After the EWSB these two scalar doublets can be written in the following form in the unitary gauge,
\begin{equation}
H=\left( 0,  \;\;  \frac{ v +h }{\sqrt 2} \right)^T , ~~~\eta=\left( \eta^\pm, \;\; \frac{\eta_R+i\eta_I}{\sqrt 2}\right)^T.
\end{equation}
The scalar potential of the model is
\begin{equation}
\begin{aligned}
V =  \mu_H^2|H|^2 +\mu_{\eta}^2|\eta|^2+\frac{\lambda_H}{2}|H|^4+\frac{\lambda_\eta}{2}|\eta|^4+\lambda_3|H|^2|\eta|^2 
 +\lambda_4|H^\dag \eta|^2 + \{\frac{\lambda_5}{2}(H^\dag \eta)^2 + \text{h.c.}\}.
\end{aligned}
\label {c}
\end{equation}
The masses of the physical scalars at tree level can be written as
\begin{eqnarray}
m_h^2 &=& \lambda_H v^2 ,\nonumber\\
m_{\eta^\pm}^2 &=& \mu_\eta^2 + \frac{1}{2}\lambda_3 v^2 , \nonumber\\
m_{\eta_R}^2 &=& \mu_\eta^2 + \frac{1}{2}(\lambda_3+\lambda_4+\lambda_5)v^2=m^2_{H^\pm}+
\frac{1}{2}\left(\lambda_4+\lambda_5\right)v^2  , \nonumber\\
m_{\eta_I}^2 &=& \mu_\eta^2 + \frac{1}{2}(\lambda_3+\lambda_4-\lambda_5)v^2=m^2_{H^\pm}+
\frac{1}{2}\left(\lambda_4-\lambda_5\right)v^2.
\label{mass_relation}
\end{eqnarray}
Here $m_h$, $m_{\eta_R}$, and $m_{\eta_I}$ are the masses of the SM like Higgs boson, the CP even and CP odd scalars from the inert doublet, respectively. $m_{\eta^\pm}$ is the mass of the charged scalar. Without any loss of generality, we consider $\lambda_5 <0, \lambda_4+\lambda_5 <0$ so that the CP even scalar is the lightest $Z_2$ odd particle and hence a stable dark matter candidate.

Denoting the squared physical masses of neutral scalar and pseudo-scalar parts of $\eta$ as $m^2_{R,I}=m^2_{\eta_R,\eta_I}$ and the mass of the right handed neutrino $N_k$ in the internal line as $M_k$, the one loop neutrino mass can be estimated as \cite{Ma:2006km}
\begin{align}
(M_{\nu})_{ij} \ & = \ \sum_k \frac{y_{ik}y_{jk} M_{k}}{32 \pi^2} \left ( \frac{m^2_{\eta_R}}{m^2_{\eta_R}-M^2_k} \: \text{ln} \frac{m^2_{\eta_R}}{M^2_k}-\frac{m^2_{\eta_I}}{m^2_{\eta_I}-M^2_k}\: \text{ln} \frac{m^2_{\eta_I}}{M^2_k} \right) \nonumber \\ 
& \ \equiv  \ \sum_k \frac{y_{ik}y_{jk} M_{k}}{32 \pi^2} \left[L_k(m^2_{\eta_R})-L_k(m^2_{\eta_I})\right] \, , \nonumber \\
\Lambda_k &= \frac{ M_{k}}{32 \pi^2} \left[L_k(m^2_{\eta_R})-L_k(m^2_{\eta_I})\right],
\label{numass2}
\end{align}
where 
$M_k$ is the mass eigenvalue of the right handed neutrino mass eigenstate $N_k$ in the internal line and the indices $i, j = 1,2,3$ run over the three neutrino generations.
The function $L_k(m^2)$ is defined as 
\begin{align}
L_k(m^2) \ = \ \frac{m^2}{m^2-M^2_k} \: \text{ln} \frac{m^2}{M^2_k} \, .
\label{eq:Lk}
\end{align}
From the physical scalar masses given above, we note that $m^2_{\eta_R}-m^2_{\eta_I}=\lambda_5 v^2$.
 In this model for the neutrino mass to match with experimentally observed limits ($\sim 0.1$~eV), Yukawa couplings of the order $10^{-3}$ are required if  $M_k$ is as low as 1 TeV and the mass difference between $\eta_R$ and $\eta_I$ is kept around 1 GeV. Such a small mass splitting between $\eta_R$ and $\eta_I$ will correspond to small quartic coupling $\lambda_5 \sim 10^{-4}$. Thus, one can suitably choose the Yukawa couplings, quartic coupling $\lambda_5$ and $M_k$ in order to arrive at sub eV light neutrino masses. 
To be in exact agreement with light neutrino masses, we first rewrite the neutrino mass given above in Eq. \eqref{numass2} in the form of type I seesaw formula 
\begin{align}
M_\nu \ = \ y \widetilde{M}^{-1} y^T \, ,
\label{eq:nu2}
\end{align}
where we have introduced the diagonal matrix $\widetilde{M}$ with elements
\begin{align}
 \widetilde{M}_i \ & = \ \frac{2\pi^2}{\lambda_5}\zeta_i\frac{2M_i}{v^2} \, , \\
\textrm {and}\quad \zeta_i & \ = \  \left(\frac{M_{i}^2}{8(m_{\eta_R}^2-m_{\eta_I}^2)}\left[L_i(m_{\eta_R}^2)-L_i(m_{\eta_I}^2) \right]\right)^{-1} \, . \label{eq:zeta}
\end{align}
The light neutrino mass matrix~\eqref{eq:nu2} is diagonalised by the usual Pontecorvo-Maki-Nakagawa-Sakata (PMNS) mixing matrix $U$, which is determined from the neutrino oscillation data (up to the Majorana phases): 
\begin{align}
D_\nu \ = \ U^\dag M_\nu U^* \ = \ \textrm{diag}(m_1,m_2,m_3) \, .
\end{align}   
Then the Yukawa coupling matrix satisfying the neutrino data can be written as
\begin{align}
y \ = \ U D_\nu^{1/2} O \widetilde{M}^{1/2} \, ,
\label{eq:Yuk}
\end{align}
where $O$ is an arbitrary complex orthogonal matrix. This is the equivalent of the Casas-Ibarra parametrisation \cite{Casas:2001sr} for scotogenic model \cite{Toma:2013zsa}.

\section{Leptogenesis from annihilations}
\label{sec:level3}
In the minimal scotogenic model discussed in the previous section, there are different types of annihilation processes which violate lepton number. They are namely,
\begin{enumerate}
    \item annihilation process of scalar doublet $\eta$: $\eta \eta \rightarrow L_\alpha L_\beta$.
    \item coannihilation process of scalar doublet and one of the singlet fermions: $\eta N \rightarrow L X$ where ($X \equiv h,\gamma,W^{\pm},Z$).
\end{enumerate}
Interestingly, if we put the additional constraints that such lepton number violating annihilations and coannihilations also generate a non-zero CP asymmetry, they lead to two different DM possibilities namely,
\begin{enumerate}
    \item the lightest neutral component of inert scalar doublet $\eta$ as DM,
    \item the lightest right handed neutrino $N$ as DM.
\end{enumerate}

The Boltzmann equations for leptonic asymmetry is given as follows:

\begin{align}
    \frac{dY_{\Delta L}}{dz} &=  \frac{1}{zH(z)}\left[\sum_i \epsilon_{N_i}(Y_{N_i} - Y^{\rm eq}_{N_i})\langle \Gamma_{N_i \rightarrow L_\alpha \eta}\rangle - Y_{\Delta L}r_{N_i}\langle\Gamma_{N_i \rightarrow L_\alpha \eta}\rangle -Y_{\Delta L}r_\eta s\langle \Gamma_{\eta \rightarrow N_1 L}\rangle\right. \nonumber \\
     &+ 2\epsilon_{\eta \eta} s\langle \sigma v\rangle_{\eta \eta \rightarrow L L}\left(Y^2_{\eta} - (Y_{\eta}^{\rm eq})^2\right)-Y_{\Delta L}Y^{\rm eq}_L r^2_\eta s\langle  \sigma v\rangle_{\eta \eta \rightarrow LL} \nonumber \\
     &+  \sum_i \epsilon_{N_i \eta} s\langle \sigma v\rangle_{\eta N_i\rightarrow L {\rm SM}}\left(Y_{\eta}Y_{N_i} - Y^{\rm eq}_{\eta}Y^{\rm eq}_{N_i}\right)- \frac{1}{2}Y_{\Delta L}Y^{\rm eq}_L r_{N_i}r_\eta s\langle \sigma v\rangle_{\eta N_i \rightarrow {\rm SM} \overline{L}} \nonumber \\
     &\left.-Y_{\Delta L}Y^{\rm eq}_\eta s\langle \sigma v \rangle^{wo}_{\eta L \rightarrow \eta \overline{L}} -\sum_iY_{\Delta L}Y^{\rm eq}_\eta s\langle \sigma v \rangle^{wo}_{\eta L \rightarrow N_iX} -\sum_iY_{\Delta L}Y^{\rm eq}_{N_i} s\langle \sigma v \rangle^{wo}_{N_i L \rightarrow \eta X} \right], \label{eq:asym} \\
     H(z) &= \sqrt{\frac{4\pi^3 g_*}{45}}\frac{M^2_{\rm DM}}{z^2M_{\rm PL}}, \quad s = g_* \frac{2\pi^2}{45}\left(\frac{M_{\rm DM}}{z}\right)^3, \nonumber \\
     r_j &= \frac{Y^{\rm eq}_j}{Y^{\rm eq}_L}, \quad  \quad
     \langle \Gamma_{j\rightarrow X} \rangle = \frac{K_1(M_j/T)}{K_2(M_j/T)}\Gamma_{j\rightarrow X}, \nonumber
\end{align}
where $z=\frac{M_{\rm DM}}{T}$, $M_{\rm PL}$ is the Planck mass and $Y=n/s$ 
denotes the comoving number density, as the ratio of number density to entropy density. The details of the derivation of this Boltzmann equation as well as the relevant equations for DM are presented in appendix \ref{sec:appen1}.
In the above equation,
$\epsilon_{\eta\eta}$ and $\epsilon_{N_i \eta}$ will be given appropriately later and
 $\epsilon_{N_i}$ is taken from \cite{Hugle:2018qbw}.

 \begin{align}
     \epsilon_{N_i} &= \sum_\alpha \frac{1}{8\pi (y^\dagger y)_{ii}}\sum_{j\neq i}\left[f\left(\frac{M^2_{N_j}}{M^2_{N_i}},\frac{m^2_\eta}{M^2_{N_i}}\right)\Im[y^*_{\alpha ,i}y_{\alpha ,j}(y^\dagger y)_{ij}] \right. \nonumber \\
     &-\left.\frac{M^2_{N_i}}{M^2_{N_j}-M^2_{N_i}}\left(1-\frac{m^2_\eta}{M^2_{N_i}}\right)^2\Im[y^*_{\alpha i}y^*_{\alpha j}H_{ij}]\right] \\
     H_{ij} &= (y^\dagger y)_{ij}\frac{M_{N_j}}{M_{N_i}} + (y^\dagger y)^*_{ij}; \quad f(x,w) = \sqrt{x}\left[1+\frac{1-2w+x}{(1-w)^2}\ln\left(\frac{x-w^2}{1-2w+x}\right)\right] \nonumber
 \end{align}
 $M_{DM}$ in $s$ is the mass of $Z_2$ odd particle taken appropriately depending on the scenario mentioned above.
Here $K_n$ is the $n$th order Modified Bessel function of second kind. The details of the Boltzmann equations are given in appendix \ref{sec:appen1}. We will now discuss this general framework in the context of two specific scenarios of DM mentioned above in the upcoming subsections.
 
The above Boltzmann equation contains the next leading order (NLO) contributions to lepton asymmetry.  For usual type I seesaw leptogenesis, such NLO effects have been calculated already, for example, see \cite{Salvio:2011sf} and references therein. It is therefore necessary to include or consider all the diagrams which can either contribute to lepton asymmetry or washout at the same order of couplings. We first only show the annihilation of dark sector particles into SM ones in Fig. \ref{fig:coasym} for the case of scalar doublet dark matter. We do not present the usual well-known two body decay diagrams (both tree and one loop) for simplicity. Based on the diagrams shown in Fig. \ref{fig:coasym}, one can construct several other diagrams just by interchanging initial and final state particles. For example, it is straightforward to consider three body decay diagram of $N_i$ into $\eta, L, X, X \equiv \gamma,  W^{\pm}, Z$. This diagram, which contributes to the asymmetry  at order $\mathcal{O}(y^4g^2)$ will  be suppressed due to not only phase space but also additional $\mathcal{O}(g^2)$ suppression. 

\begin{figure}
      \centering
  \begin{tabular}{lr}
      $\mathcal{O}(Y^4g^2)$:\\
         \begin{tikzpicture}[/tikzfeynman/small]
         \begin{feynman}
         \vertex (i){$\eta$};
         \vertex [right = 4.cm of i] (j){$X$};
         \vertex [below = 2.cm of i] (k){$N_i$};
         \vertex [below = 2cm of j] (l){$L_\alpha$};
         \vertex [right = 1.cm of k] (k1);
         \vertex [left = 1.cm of l] (l1);
         \vertex [right = 2.cm of i] (v1);
         \vertex [below = 2.cm of v1] (v2);
         \vertex [below = 1.cm of v1] (v3);
         \diagram*[small]{(i) -- [charged scalar] (v1),(v1) -- [charged scalar,edge label = $\eta$] (v3),(k) -- [ fermion] (k1),(k1) -- [anti fermion,edge label=$L_\beta$] (v3),(k1) -- [charged scalar,edge label'=$\eta$] (l1),(v3) -- [ majorana,edge label=$N_j$] (l1),(v1) -- [boson] (j),(l1) -- [fermion] (l)};
         \end{feynman}
         \end{tikzpicture}
         &
         \begin{tikzpicture}[/tikzfeynman/small]
         \begin{feynman}
         \vertex (i){$\eta$};
         \vertex [below = 1.6cm of i] (j){$N_i$};
         \vertex [right = 1.cm of i] (l1);
         \vertex [below = 1.6cm of l1] (l3);
         \vertex [right = 1.cm of l1] (l2);
         \vertex [below = 0.8cm of l2] (v1);
         \vertex [right = 4.2cm of i] (k){$X$};
         \vertex [right = 1.2cm of v1] (v2);
         \vertex [below = 1.6cm of k] (m){$L_\alpha$};
         \diagram*[small]{(i) -- [charged scalar] (l1),(l1) -- [majorana,edge label=$N_j$](v1),(v1)--[anti charged scalar,edge label = $\eta$](l3),(l3)--[anti fermion,edge label = $L_\beta$](l1),
(j)--[fermion](l3),(v1) -- [fermion,edge label = $L_\alpha$](v2),(v2)--[boson](k),(v2)--[fermion](m)};
\end{feynman}
\end{tikzpicture}
\\
\begin{tikzpicture}[/tikzfeynman/small]
\begin{feynman}
\vertex (i){$N_i$};
\vertex[right = 1.cm of i] (v1);
\vertex[above right = 1.cm of v1] (v2);
\vertex[above right = 1.cm of v2] (j){$\eta$};
\vertex[below right = 1.cm of v2] (v3);
\vertex[right = 1.cm of v3] (v4);
\vertex[above right = 1.cm of v4](k){$L_\alpha$};
\vertex[right = 1.cm of v4](l){$X$};
\diagram*[small]{(i) -- [fermion](v1)--[anti fermion,edge label=$L_\beta$](v2)--[majorana,edge label=$N_j$](v3)--[fermion,edge label=$L_\alpha$](v4)--[fermion](k),(v2)--[anti charged scalar](j),(v3)--[anti charged scalar,edge label=$\eta$](v1),(v4)--[boson](l)};
\end{feynman}
\end{tikzpicture}
&
\begin{tikzpicture}[/tikzfeynman/small]
\begin{feynman}
\vertex (i){$N_i$};
\vertex[right = 1.cm of i] (v1);
\vertex[above right = 1.cm of v1] (v2);
\vertex[above right = 1.cm of v2] (j){$L_\alpha$};
\vertex[below right = 1.cm of v2] (v3);
\vertex[right = 1.cm of v3] (v4);
\vertex[above right = 1.cm of v4](k){$\eta$};
\vertex[right = 1.cm of v4](l){$X$};
\diagram*[small]{(i)--[fermion](v1)--[anti fermion,edge label'=$L_\beta$](v3)--[majorana,edge label'=$N_j$](v2)--[fermion](j),(v1)--[charged scalar,edge label=$\eta$](v2),(v3)--[anti charged scalar](v4)--[anti charged scalar](k),(v4)--[boson](l)};
\end{feynman}
\end{tikzpicture} \\
\begin{tikzpicture}[/tikzfeynman/small]
\begin{feynman}
\vertex (i){$\eta$};
\vertex [below right= 1.414cm of i] (v1);
\vertex [below left= 2.cm of v1] (j){$N_i$};
\vertex [above right = 1.cm of j] (i1);
\vertex [above right = 0.5cm of i1] (i2);
\vertex [right = 1.cm of v1] (v2);
\vertex [right = 3.cm of i] (m){$X$};
\vertex [below = 2.cm of m] (o){$L_\alpha$};
\diagram*[small]{(i) -- [charged scalar] (v1),(v1) -- [fermion,edge label = $L_\alpha$] (v2),(v2) -- [ boson] (m),(v2) -- [fermion] (o),(j) -- [fermion] (i1)--[anti fermion,half right,edge label'=$L_\beta$](i2)--[majorana,edge label'=$N_j$](v1),(i2)--[anti charged scalar,half right,edge label'=$\eta$](i1)};
\end{feynman}
\end{tikzpicture}
& 
\begin{tikzpicture}[/tikzfeynman/small]
\begin{feynman}
\vertex (i){$\eta$};
\vertex [right = 3.cm of i] (j){$X$};
\vertex [below = 1.5cm of i] (i1);
\vertex [left = 1.2cm of i1] (k){$N_i$};
\vertex [right = 1.5cm of k] (i2);
\vertex [right = 0.5cm of i2] (i3);
\vertex [below = 1.5cm of j] (l){$L_\alpha$};
\vertex [right = 1.5cm of i] (v1);
\vertex [below = 1.5cm of v1] (v2);
\diagram*[small]{(i) -- [charged scalar] (v1),(v1) -- [charged scalar,edge label = $\eta$] (v2),(v1) -- [boson] (j),(v2) -- [ fermion] (l),(k) -- [fermion] (i2),(i2)--[anti fermion,half right,edge label'=$L_\beta$](i3)--[majorana,edge label'=$N_j$](v2),(i3)--[charged scalar,half right,edge label'=$\eta$](i2)};
\end{feynman}
\end{tikzpicture}\\
      $\mathcal{O}(Y^6)$:\\
       \begin{tikzpicture}[/tikzfeynman/small]
        \begin{feynman}
        \vertex (i){$\eta$};
        \vertex [below = 1.6cm of i] (j){$\eta$};
        \vertex [right = 1.cm of i] (l1);
        \vertex [right = 1.cm of l1] (v1);
        \vertex [right = 1.cm of v1] (l2);
        \vertex [below = 0.8cm of v1] (l3);
        \vertex [below = 1.6cm of v1] (v2);
        \vertex [right = 0.8cm of l2] (k){$L_\alpha$};
        \vertex [right = 1.6cm of v2] (m){$L_\beta$};
        \diagram*[small]{(i) -- [charged scalar] (l1),
        (l1) -- [ majorana,edge label = $N_1$] (l2),
        (l2) -- [anti charged scalar, edge label = $\eta$] (l3),
        (l3) -- [anti fermion,edge label = $L_\gamma$] (l1),
        (l3) -- [anti fermion,edge label = $N_j$] (v2),
        (l2) -- [fermion] (k),
        (j) -- [charged scalar] (v2),
        (v2) -- [fermion](m)};
        \end{feynman}
    \end{tikzpicture}
    &
    \begin{tikzpicture}[/tikzfeynman/small]
        \begin{feynman}
        \vertex (i){$\eta$};
        \vertex [below = 1.6cm of i] (j){$\eta$};
        \vertex [right = 1.cm of j] (l1);
        \vertex [right = 1.cm of l1] (v1);
        \vertex [right = 1.cm of v1] (l2);
        \vertex [above = 0.8cm of v1] (l3);
        \vertex [above = 1.6cm of v1] (v2);
        \vertex [right = 0.8cm of l2] (k){$L_\beta$};
        \vertex [right = 1.6cm of v2] (m){$L_\alpha$};
        \diagram*[small]{(j) -- [charged scalar] (l1),
        (l1) -- [ majorana,edge label' = $N_1$] (l2),
        (l2) -- [anti charged scalar,edge label' = $\eta$] (l3),
        (l3) -- [anti fermion,edge label' = $L_\gamma$] (l1),
        (l3) -- [anti fermion,edge label' = $N_j$] (v2),
        (l2) -- [fermion] (k),
        (i) -- [charged scalar] (v2),
        (v2) -- [fermion](m)};
        \end{feynman}
    \end{tikzpicture}
     \end{tabular}
\caption{1-loop diagrams contributing to the asymmetry arising from the interference at $\mathcal{O}(y^4g^2)$ and $\mathcal{O}(y^6)$.}
      \label{fig:loop}
      \label{fig:coasym2}
  \end{figure}
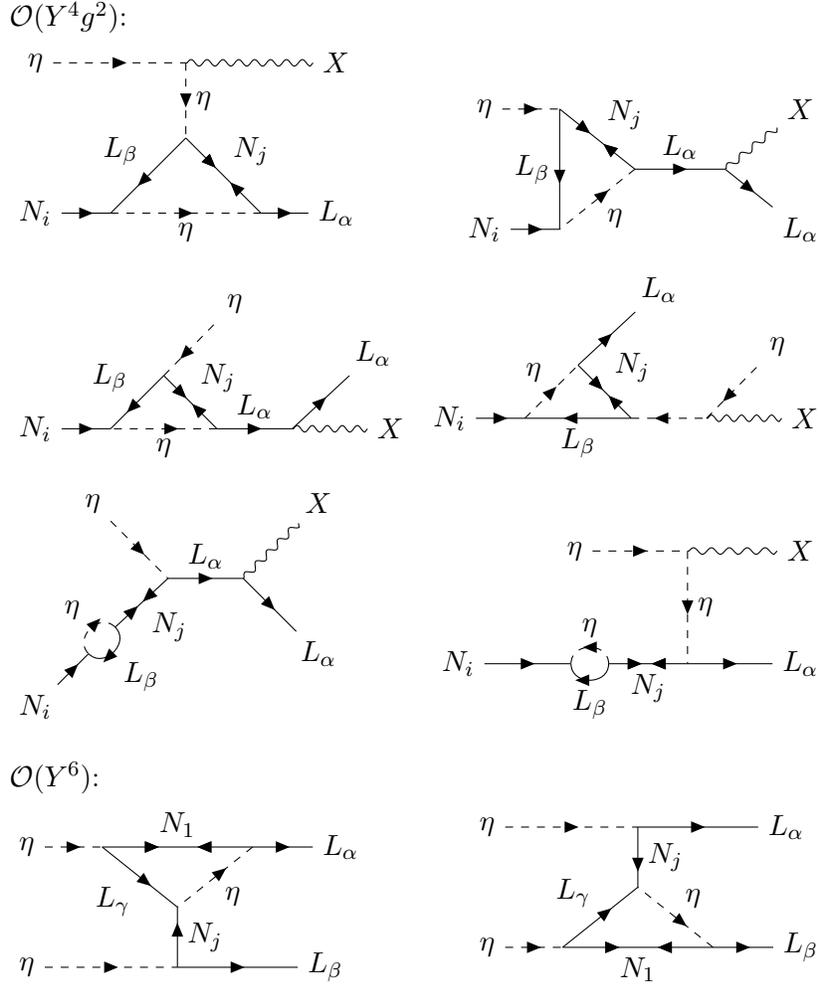

Now, although the 3-body decay and the co-annihilation are of the same order,  the contribution in the Boltzmann evolution is different as they enter the Boltzmann equation differently. The 3-body decay will enter into the equation as an addition to the 2-body decay in which the contribution coming from the 2 body decay will dominate. As will be shonw later, the contributions of coannihilation to lepton asymmetry may dominant over that from decay process. This is due to fact that the imaginary part of the interference is not suppressed as it becomes two $2\rightarrow2$ processes (referring to the diagrams in the second line in Fig. \ref{fig:loop} leading to the interference at order $\mathcal{O}(y^4g^2)$) both mediated by the leptons after cutting loop diagrams, whereas in case of the decay the loop diagram becomes $1\rightarrow2$ and $2\rightarrow2$ in which the $2\rightarrow2$ is a $t$-channel process mediated by the right-handed neutrino after cutting. 
We present all such 1-loop diagrams contributing to the asymmetry arising from the interference at order $\mathcal{O}(y^4g^2)$ and $\mathcal{O}(y^6)$  in Fig. \ref{fig:loop}. Similarly, there exists several washout processes that can be constructed by swapping initial and final state particles.  This is true for fermion dark matter as well, which we discuss in one of the upcoming sections. To point out the significance of such processes at the same NLO, we therefore make a comparison of different washout processes (both inverse decay and scatterings) and show them in in Fig. \ref{fig:LLS} (lightest scalar as Dark Matter) and in Fig. \ref{fig:LLF} (lightest fermion as Dark Matter).
 \begin{figure}
\centering
\begin{tabular}{lr}
\begin{tikzpicture}[/tikzfeynman/small]
\begin{feynman}
\vertex (i){$\eta$};
\vertex [below = 2.cm of i] (j){$N_i$};
\vertex [below right= 1.414cm of i] (v1);
\vertex [right = 1.cm of v1] (v2);
\vertex [right = 3.cm of i] (m){$X$};
\vertex [below = 2.cm of m] (o){$L_\alpha$};
\diagram*[small]{(i) -- [charged scalar] (v1),(v1) -- [fermion,edge label = $L_\alpha$] (v2),(v2) -- [ boson] (m),(v2) -- [fermion] (o),(j) -- [anti fermion] (v1)};
\end{feynman}
\end{tikzpicture}
& 
\begin{tikzpicture}[/tikzfeynman/small]
\begin{feynman}
\vertex (i){$\eta$};
\vertex [right = 3.cm of i] (j){$X$};
\vertex [below = 1.5cm of i] (k){$N_i$};
\vertex [below = 1.5cm of j] (l){$L_\alpha$};
\vertex [right = 1.5cm of i] (v1);
\vertex [below = 1.5cm of v1] (v2);
\diagram*[small]{(i) -- [charged scalar] (v1),(v1) -- [charged scalar,edge label = $\eta$] (v2),(v1) -- [boson] (j),(v2) -- [ fermion] (l),(k) -- [anti fermion] (v2)};
\end{feynman}
\end{tikzpicture}
\end{tabular}      
\caption{Feynman diagrams contributing to $\langle\sigma v\rangle_{{\rm DM DM} \rightarrow X L}$ and the asymmetry $\epsilon$ at leading order. Here $X \equiv \gamma,W^{\pm},Z$.}
\label{fig:coasym}
\end{figure}
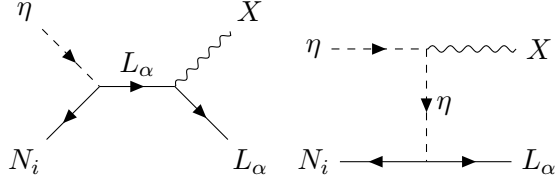

\subsection{Scalar doublet $\eta$ as Dark Matter}

In this scenario the only way to get asymmetry is through co-annihilations. Pure scalar annihilations give rise to vanishing leptonic asymmetry if $\eta$ is the lightest $Z_2$ odd particle, as required for it to be the DM candidate. This is particularly due to the fact the "on-shell" criteria of loop particles can not be realised in such case, resulting in a vanishing CP asymmetry, as we discuss below.

The relevant coannihilation processes, both tree level as well as one loop level, are shown in Fig.\ref{fig:coasym} and in Fig.\ref{fig:coasym2}. One may notice that the one loop self-energy diagrams, arising from the lepton propagator, do not  contribute to the lepton asymmetry because the processes occur before electroweak symmetry breaking and thus lepton mass should be zero giving rise to vanishing self-energy loop contribution. 
Thus, only interference between tree and one loop vertex correction can give rise to CP asymmetry.
For this scenario the Boltzmann equations 
for the $Z_2$ odd particles take the following form:
\begin{align}
    \frac{dY_{N_k}}{dz} &= -\frac{1}{zH(z)}\left[(Y_{N_k}-Y^{\rm eq}_{N_k})\langle \Gamma_{N_k \rightarrow L_\alpha \eta}\rangle + (Y_{N_k}Y_{\eta}-Y^{\rm eq}_{N_k}Y^{\rm eq}_{\eta})s\langle \sigma v \rangle_{\eta N_k\rightarrow L{\rm SM}} \right. \nonumber \\
    &+ \left. \sum_{l=1}^3(Y_{N_k}Y_{N_l}-Y^{\rm eq}_{N_k}Y^{\rm eq}_{N_l})s\langle \sigma v \rangle_{N_l N_k\rightarrow {\rm SM SM}}\right], \nonumber \\
    \frac{dY_{\eta}}{dz} &= \frac{1}{zH(z)}\left[ 
    (Y_{N_k}-Y^{\rm eq}_{N_k})\langle \Gamma_{N_i \rightarrow L_\alpha \eta}\rangle 
     - 2(Y^2_{\eta}-(Y^{\rm eq}_{\eta})^2)s\langle \sigma v\rangle_{\eta \eta \rightarrow {\rm SM SM}}\right. \nonumber \\
    &- \left. \sum^3_{m=1}(Y_{N_m}Y_{\eta}-Y^{\rm eq}_{N_m}Y^{\rm eq}_{\eta})s\langle \sigma v \rangle_{\eta N_m\rightarrow L{\rm SM}} \right].
\end{align}

The CP asymmetry arising from the interference between tree and 1-loop diagrams in Fig. \ref{fig:coasym} and Fig. \ref{fig:coasym2} can be estimated as
\begin{align}
\epsilon_{N_i \eta} &= \frac{2}{(yy^\dagger)_{ii}} \sum_{j}\Im[(yy^\dagger)^2_{ij}]\widetilde{\epsilon}_{ij}, \label{eq:asymB} \\
 \widetilde{\epsilon}_{ij}   &= \frac{\sqrt{x_j}}{6 x_i
   \left(-x_i^{3/2}+x_i (x_j-2)+\sqrt{x_i} x_j+1\right)^2(\sqrt{x_i}-3)} \left(x_i^{7/2} (3 x_j+1) +\sqrt{x_i} (3 x_j+5)+1 \right.\nonumber \\
   &- \left.3 x_i^{5/2} \left(x_j \left(D+(x_j-3) x_j+4\right)  -  3 D -2\right)-3 x_i^{3/2}
   \left(2 \left(D+3\right) + x_j \left(x_j \left(D+x_j+1\right)-D-4\right)\right) \right. \nonumber \\
   &- \left. x_i^4+f^3 \left(3
   D+3 x_j^2+11\right) - 3 x_i^2 \left(x_j \left(D+2 (x_j-1) x_j+2\right) - D +6\right)+x_i \left(1-3
   x_j \left(D+x_j-4\right)\right) \right) \nonumber \\
   &+ \frac{\sqrt{x_j}}{4 x_i}\left(\sqrt{x_i}-1+\frac{\sqrt{x_j}}{(1+\sqrt{x_i})^2}(\sqrt{x_i}-1+r_j)\left(\log\left(\frac{1+\sqrt{x_i}x_j}{x_i(1+\sqrt{x_i})}\right)-\log\left(\frac{1+x_i+x^{3/2}_i+\sqrt{x_i}x_j}{x_i(1+\sqrt{x_i})} \right)\right.\right. \nonumber \\
   &+ \left.\left. \log\left(1+\frac{1+\sqrt{x_i}}{\sqrt{x_i}(\sqrt{x_i}-1+x_i+x_j)}\right)\right)\right) + \frac{\sqrt{x_i} x_j\tilde{\Gamma}_j}{\pi((x_i-x_j)^2 + x_j\tilde{\Gamma}^2_j)} \label{eq:coanneps} \\
   D &= \sqrt{(x_i-x_j) \left(x_i+4 \sqrt{x_i}-x_j+4\right)} \qquad
  x_l = \frac{M^2_{N_l}}{m^2_\eta} \qquad \tilde{\Gamma}_j = \frac{\Gamma_j}{m_\eta}. \nonumber
\end{align}
where the details of the asymmetry is shown in appendix \ref{sec:appen3}. Although the above expression is an $s-$ wave approximation for actual expression shown in appendix \ref{sec:appen3}, we have used the actual expression for our analysis.   It should be noted that in the above expression always ($1\leq x_j\leq x_i$) where $j$ stands for $N_j$ inside the loop while $i$ stands for $N_i$ as one of the initial state particles, shown in Fig. \ref{fig:coasym} and in Fig.\ref{fig:coasym2}. This is simply to realise the "on-shell" -ness of the loop particles in order to generate the required CP asymmetry. 

There are several wash-out processes in this scenario, categorised as follows:
\begin{itemize}
\item $\bf \Delta L = 2 $: $L \eta \rightarrow \overline{L} \eta, \eta \eta \rightarrow L L$ are purely wash-out processes.
\item $\bf \Delta L = 1 $: there are two main sources of such wash-out, namely
\begin{enumerate}
\item inverse decay of $\Gamma(N_k \rightarrow L \eta)$,
\item inverse process of co-annihilation $N_k \eta \rightarrow L,X(=\gamma, W, Z, h)$.
\end{enumerate}
\end{itemize}
We have taken them into account in our numerical calculations.

Adopting the Casas-Ibarra parametrisation given in Eq. \eqref{eq:Yuk} ,
we see that CP phases in $U$ do not contribute to $\epsilon_{N_i\eta}$, but
complex variables in the orthogonal matrix $O$ can lead to non-vanishing value of $\epsilon_{N_i\eta}$. This is similar to leptogenesis from pure decay in this model \cite{Hugle:2018qbw} where, in the absence of flavour effects, the orthogonal matrix $O$ played a crucial role. 
In general, this $3\times3$ orthogonal matrix $O$ can be parametrised by three complex parameters of type $\theta_{\alpha \beta} = \theta^R_{\alpha \beta} + i\theta^I_{\alpha \beta}, \theta^R_{\alpha \beta} \in [0, 2\pi], \theta^I_{\alpha \beta} \in \mathbb{R}$ \cite{Ibarra:2003up} \footnote{For some more discussions on different possible structure of this matrix and implications on a particular leptogenesis scenario in this model, we refer to the recent work \cite{Mahanta:2019gfe}.}.
In general, the orthogonal matrix $O$ for $n$ flavours can be product of $^nC_2$ number of rotation matrices of type
\begin{align}
    O_{\alpha \beta} &= \begin{pmatrix} \cos{(\theta^R_{\alpha \beta} + i\theta^I_{\alpha \beta})} & \cdots & \sin{(\theta^R_{\alpha \beta} + i\theta^I_{\alpha \beta})} \\
    \vdots & \ddots & \vdots \\
  - \sin{(\theta^R_{\alpha \beta} + i\theta^I_{\alpha \beta})} & \cdots & \cos{(\theta^R_{\alpha \beta} + i\theta^I_{\alpha \beta})} \end{pmatrix},
\end{align}
with rotation in the $\alpha-\beta$ plane and dots stand for zero. For example, taking $\alpha=1, \beta=2$ we have
\begin{align}
  O_{12} &= \begin{pmatrix} \cos{(\theta^R_{12} + i\theta^I_{12})} &  \sin{(\theta^R_{12} + i\theta^I_{12})} & 0 \\
  -\sin{(\theta^R_{12} + i\theta^I_{12})} &  \cos{(\theta^R_{12} + i\theta^I_{12})} & 0 \\
  0 &  0 & 1\end{pmatrix}.
\end{align}
The above asymmetry along with this rotation (one at a time) takes the following form 

\begin{align}
    \epsilon_{N_i\eta} &= \sum_j\frac{(m^4_j-m^4_i)\sin{(2\theta^R_{ij})}\sinh{(2\theta^I_{ij})}}{\Lambda_j^2((m^2_i-m^2_j)\cos(2\theta^R_{ij})+(m^2_i+m^2_j)\cosh(2\theta^I_{ij}))}\widetilde{\epsilon}_{ij},
\end{align}
where $m_i$'s are the light neutrino masses and $\Lambda_i$'s are defined above in Eq. \eqref{numass2}. On the right hand side of the above equation, a summation over index $j$ is implicit.
\begin{table}[]
    \centering
    \begin{tabular}{|c|c|c|c|}
    \hline 
    Parameter & Required parameter for Correct Relic & Particle & Mass\\
     \hline   
      $\mu_\eta$  &  870 GeV & $m_{\eta_R}$ & 870 GeV\\
      $\lambda_1$ & 0.253 & $m_{\eta_I}$ & 870 GeV\\
        $\lambda_3$ & 0.65 & $m_{\eta_\pm}$ & 881 GeV\\
        $\lambda_4$ & -0.65 & $M_{N_1}$ & 1. TeV \\
        $\lambda_5$ & $8\times 10^{-6}$ & $M_{N_2}$ & 1.5 TeV\\
        $\lambda_2$ & 1 & $M_{N_3}$ & 2. TeV\\
        $\theta^R_{23}=\theta^R_{12}$ & $\frac{\pi}{4}\omega$ & &\\
        $\theta^R_{13}$ & $-\frac{\pi}{4}\omega$ & & \\
        $\theta^I_{12}$ & $\frac{3\pi}{4}\omega$ & &\\
        $\theta^I_{13}=\theta^I_{23}$ & $\frac{\pi}{4}\omega$ & &\\
        \hline
    \end{tabular}
    \caption{The numerical values of the parameters chosen for generating correct scalar DM relic and baryon asymmetry. We denote it as benchmark point 1 (BP1). The fraction $\omega$ is to ensure to get the correct baryonic asymmetry. For Normal (Inverted) Hierarchy $\omega = 0.7(0.56)$.}
    \label{tab:BPS}
\end{table}
As an example, we have taken the benchmark values shown in table \ref{tab:BPS} to compute the baryon asymmetry as well as scalar DM relic numerically.
Here we consider 
$\eta_I$ as the DM candidate ($\eta_I \equiv \rm DM$, corresponding to positive value of quartic coupling $\lambda_5$) which is similar to the inert doublet model discussed extensively in the literature \cite{Ma:2006km,Barbieri:2006dq, LopezHonorez:2006gr}. Typically there exists two distinct mass regions, $M_{\rm DM} \leq 80$ GeV and  $M_{\rm DM} \geq 500$ GeV, where correct relic abundance criteria can be satisfied. In both regions, depending on the mass differences $m_{\eta^\pm}-m_{\eta_I}, m_{\eta_R}-m_{\eta_I}$, the coannihilations of $\eta_I, \eta^\pm$ and $\eta_R, \eta_I$ can also contribute to the DM  relic abundance \cite{Griest:1990kh, Edsjo:1997bg}. 
As for the mixing angles in the PMNS matrix $U$ we took the best fit values obtained from the recent global fit analysis \cite{Esteban:2018azc} shown in the table \ref{tab:pmns}.
\begin{table}[!h]
\begin{tabular}{|c|c|c|c|c|c|}
\hline
$\theta_{12}$ & $\theta_{13}$ & $\theta_{23}$ & $\Delta m^2_{21}\times10^{-5} (\text{eV}^2)$  & $ \lvert \Delta m^2_{31} \rvert \times10^{-3} (\text{eV}^2) $ & $m_{lightest} (\textrm{eV})$\\ \hline
33.7$^\circ$ & 8.8$^\circ$ & 41.4$^\circ$ & 7.54 & 2.43 & 0.01\\ \hline
\end{tabular}
\caption{The numerical values of light neutrino parameters used in the calculations.}
\label{tab:pmns}
\end{table}

\begin{figure}[!ht]
    \centering
    \begin{tabular}{lr}
    \includegraphics[width=0.5\textwidth]{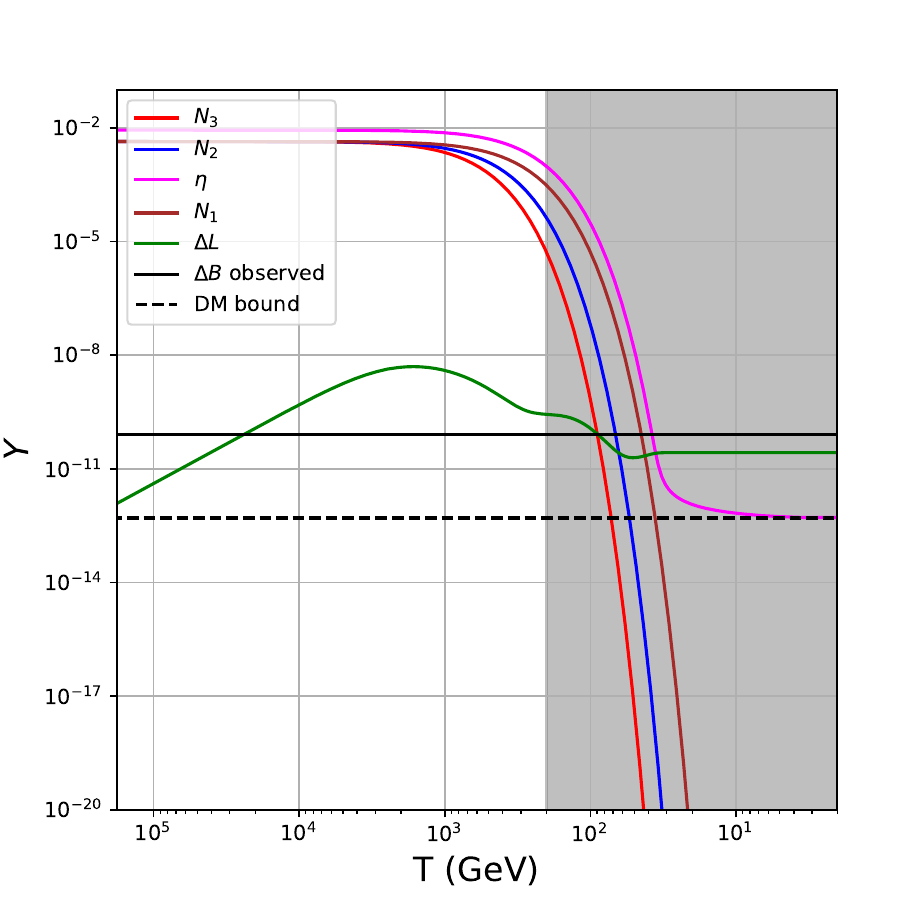}&
    \includegraphics[width=0.5\textwidth]{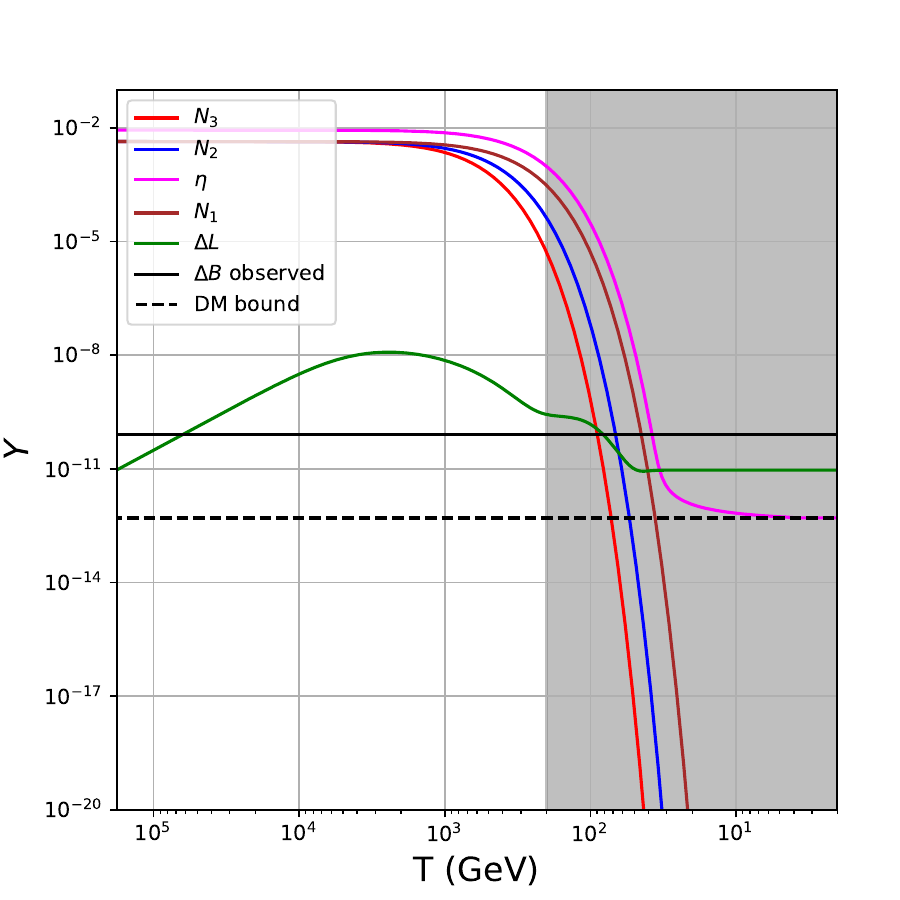} \\ 
    \includegraphics[width=0.5\textwidth]{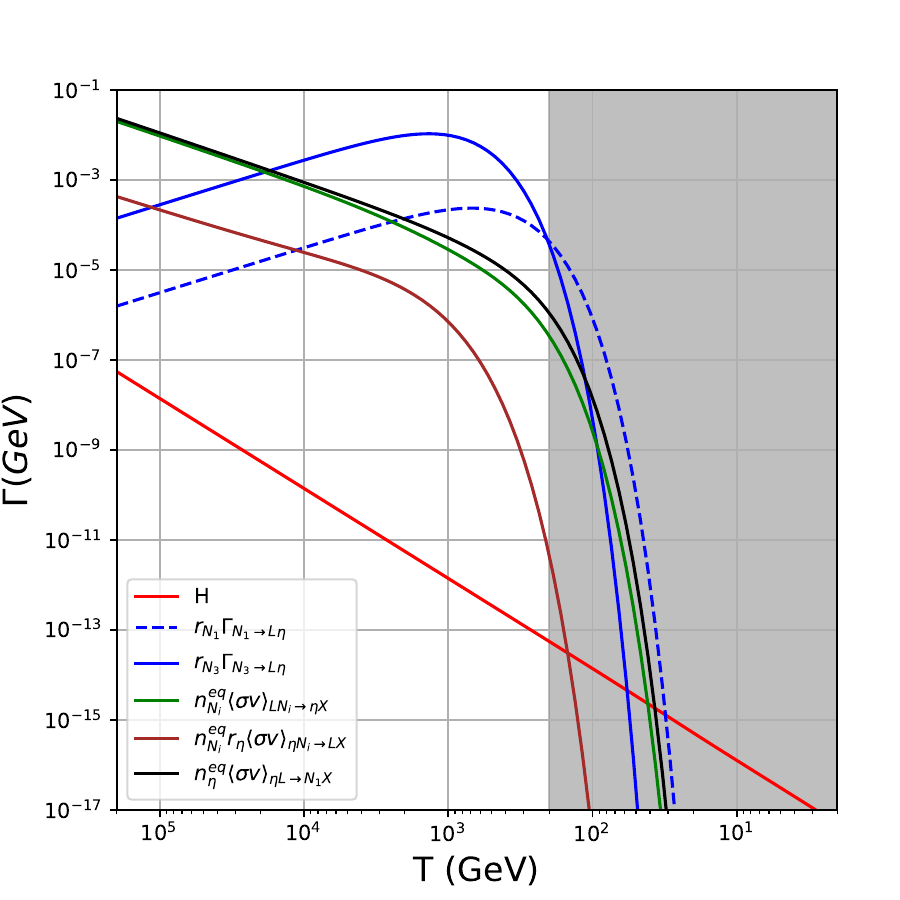}&
    \includegraphics[width=0.5\textwidth]{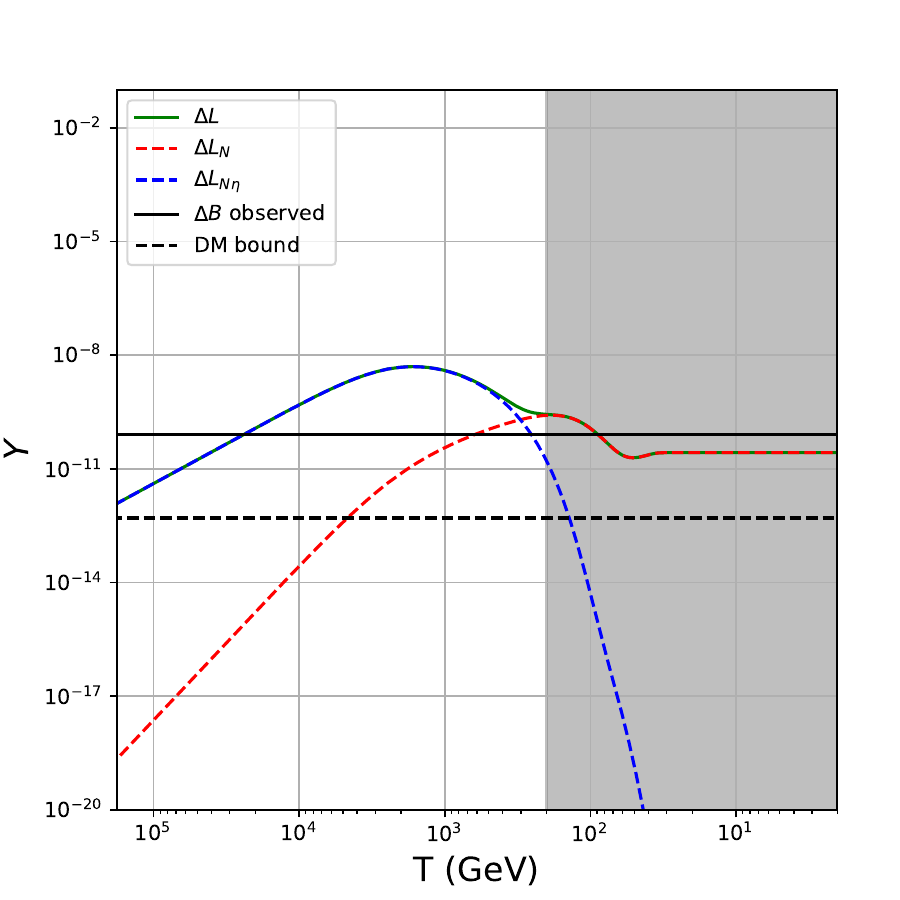}
    \end{tabular}
    \caption{Top panel: comoving number densities  of $Z_2$ odd particles and lepton asymmetry as a function of temperature for scalar DM scenario. The left (right) panel corresponds to the normal (inverted) hierarchy of neutrino mass spectrum. Bottom panel: rates of individual wash-out processes along with the Hubble rate (left panel); individual (co)annihilation contributions of each channel to the asymmetry. The solid (dashed) black line corresponds to the baryon asymmetry (DM abundance) observed at present epoch ($T \sim 0$). Shaded regions represent the epochs after sphaleron freeze out.}
    \label{fig:LLS}
\end{figure}

To perform the numerical analysis, we implement the model in \texttt{SARAH 4} \cite{Staub:2013tta} and extract the thermally averaged annihilation rates from \texttt{micrOMEGAs 4.3} \cite{Barducci:2016pcb} to use while solving the Boltzmann equations above.
In Fig. \ref{fig:LLS}, we plot the comoving number densities of all $Z_2$ odd particles, taking part in generating the lepton asymmetry along with the generated asymmetry $\Delta L$, as functions of temperature. 
The left panel corresponds to normal hierachy (NH) of neutrino mass spectrum and the right panel to the case of inverted hierarchy (IH).
The horizontal solid black line labelled as "$\Delta L$ observed" correspond to the value of $\Delta L$ that is partially converted into the observed baryon asymmetry via the electroweak sphaleron processes with the conversion factor $C_s = \frac{8N_f + 4N_H}{22N_f+13N_H}$ where $N_f=3, N_H=2$ are the number of fermion generations and Higgs doublets respectively \cite{sph1}. 
While sphalerons violate $B+L$, they conserve $B-L$ symmetry.
The sphaleron processes are effective  with a thermal rate $\Gamma_{\rm sph}\sim (\alpha_2 T)^4$ with $SU(2)$ gauge coupling constant $\alpha_2$ at high temperature until EW phase transition, while they are exponentially suppressed due to finite gauge boson masses after EW gauge symmetry breaking.
The shaded regions in the panels correspond to the temperature below
which the sphaleron processes become inoperative($T\lesssim 200$ GeV \cite{sph2}). The benchmark parameters are chosen in such a way that the generated lepton asymmetry $\Delta L$ by the epoch of sphaleron freeze-out is sufficient enough to produce the observed baryon asymmetry.
The dashed horizontal black line corresponds to the observed DM relic abundance in the present universe \cite{Aghanim:2018eyx}. As can be seen from this plot, the lepton asymmetry grows as the temperature cools down due to the contributions from the co-annihilation diagrams. While the lepton asymmetry gets converted into the baryon asymmetry at EW phase transition temperature, it takes a while for DM to freeze-out. Since non-zero CP asymmetry arises from coannihilations between $\eta$ and heavier right handed neutrinos $N_{2,3}$, the lepton number generating processes get frozen out much earlier compared to DM self annihilations. This makes sure that a net lepton asymmetry is created with the right amount without being washed out entirely.
The limits on the parameter $\lambda_5$, keeping all the other parameters same as shown on table \ref{tab:BPS}, are set by two constraints. One is the mass difference between the neutral scalar and pseudo-scalar ($\Delta = m_{\eta_R}-m_{\eta_I}$) which needs to be more than approximately $100$ keV, in order to avoid $Z$ mediated inelastic direct detection scattering of DM off nucleons, as we discuss below. The other constraint is coming from the required leptonic asymmetry with maximal CP asymmetry. The range is shown in the table \ref{tab:lambda5S}.
\begin{table}
    \centering
    \begin{tabular}{|l|c|}
    \hline
     & $\lambda_5$ \\ \hline
         NH & $(0.3-8.1)\times 10^{-5}$ \\ \hline
         IH & $(0.03-1.2)\times 10^{-4}$ \\
         \hline
    \end{tabular}
    \caption{The range of $\lambda_5$ allowed by phenomenological requirements of satisfying direct detection bounds and generating the required lepton asymmetry.}
    \label{tab:lambda5S}
\end{table}

\begin{figure}[!ht]
\centering
\begin{tabular}{c}

\begin{tikzpicture}[/tikzfeynman/small]
\begin{feynman}
\vertex (i){$\eta^0_{I}$};
\vertex [right = 1.6cm of i] (v1);
\vertex [right = 1.6cm of v1] (j){$\eta^0_{R}$};
\vertex [below = 1.6cm of i] (k){$p,n$};
\vertex [right = 1.6cm of k] (v2);
\vertex [right = 1.6cm of v2] (l){$p,n$};
\diagram*[small]{(i) --[scalar] (v1),(v1) -- [scalar](j),(v1) -- [boson, edge label=$Z$](v2),(k)--[fermion](v2),(v2)--[fermion](l)};
\end{feynman}
\end{tikzpicture}
\end{tabular}
\caption{Feynman diagram for inelastic scattering of scalar DM off nucleon mediated by Z boson. }
\label{fig:DD}
\end{figure}
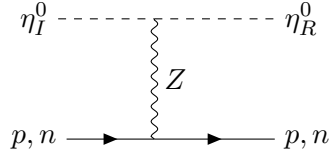

Since the DM in this scenario has electroweak gauge interactions, we include DM direct detection constraints arising from tree level $Z$ boson mediated processes $\eta_R n \rightarrow \eta_I n$, $n$ being a nucleon (shown in Fig. \ref{fig:DD}). 
The cross section for $Z$ mediated inelastic process is \cite{Cui:2009xq}
\begin{align}
\sigma^0_n &= \frac{G^2_F}{2\pi}\mu^2_n \simeq 7.44\times 10^{-39} \textrm{cm}^2.     
\end{align}
We note that one can forbid such scattering if $\delta = m_{\eta_I} - m_{\eta_R} > 100$ keV. Using the expressions for physical masses above, it leads to a lower bound on the dimensionless quartic coupling $\lambda_5$ as 
\begin{equation}
\lambda_5 \approx 1.65 \times 10^{-7} \left( \frac{\delta}{100 \; \rm keV} \right) \left( \frac{M_{\rm DM}}{100 \; \rm GeV} \right). \nonumber
\end{equation}
This lower limit on $\lambda_5$ becomes weaker for heavier DM masses. Now, if we avoid this bound i.e the mass difference between the scalar and pseudo-scalar dark matter is above 100 keV then the direct detection is majorly dominated by the first process shown in Fig. \ref{fig:DD}. Our benchmark point satisfies this constraint, forbidding such inelastic DM scattering.
The Yukawa couplings for all three generation of leptons for the chosen benchmark point (considering only 13 rotation in $O$ matrix, however)  in this scenario can be written in terms of the following matrix.
\begin{align}
 y \rvert_{NH} &= 
 \begin{pmatrix}
4.47 - 5.04i	& 10.7 + 4.77 i	& 6.35 + 7.96i \\
-0.66 - 5.30i	& 8.38 - 4.06i & 7.29 + 0.59i \\	
-6.34 - 2.74i	& 1.15 - 11.7i	& 6.20 - 8.2i
\end{pmatrix}\times 10^{-3}, \\
 y \rvert_{IH} &= \begin{pmatrix}
8.52 - 13.8i &	5.78 + 0.25i &	-5.66 - 0.20i\\	
-4.89 - 12.8i &	4.54 - 3.93i &	-4.42 + 4.03i \\	
-17.18 - 3.51i & -2.6 - 6.78i &	2.71 + 6.60i	
\end{pmatrix}\times 10^{-3}.
\end{align}
Now, we move onto the discussion of fermion singlet DM where the direct detection constraints are less severe due to its gauge singlet nature. This is the topic of our next subsection.

\subsection{Right handed neutrino as dark matter}

If the lightest $Z_2$ odd particle is the lightest of the right handed neutrinos (and hence the DM candidate), then the annihilation processes responsible for creating a non-zero lepton asymmetry are shown in Fig. \ref{fig:LL}. Once again, pure self annihilation of DM can not provide the asymmetry due to the absence of "on-shell" condition for loop particles. If the scalar doublet $\eta$ is the next to the lightest $Z_2$ odd particle, then the annihilation processes shown in Fig. \ref{fig:LL} can produce the required lepton asymmetry. For this scenario the Boltzmann equations 
for the $Z_2$ odd particles take the following form:
\begin{align}
    \frac{dY_{N_k}}{dz} &= -\frac{1}{zH(z)}\left[(Y_{N_k}-Y^{\rm eq}_{N_k})\langle \Gamma_{N_k \rightarrow L_\alpha \eta}\rangle + (Y_{N_k}Y_{\eta}-Y^{\rm eq}_{N_k}Y^{\rm eq}_{\eta})s\langle \sigma v \rangle_{\eta N_k\rightarrow L{\rm SM}} \right. \nonumber \\
    &+ \left. \sum_{l=1}^3(Y_{N_k}Y_{N_l}-Y^{\rm eq}_{N_k}Y^{\rm eq}_{N_l})s\langle \sigma v \rangle_{N_l N_k\rightarrow {\rm  SM SM}}\right], \quad \textrm{ for}~ k = 2, 3 \nonumber \\
    \frac{dY_{\eta}}{dz} &= \frac{1}{zH(z)}\left[ 
    \sum^3_{k=2}(Y_{N_k}-Y^{\rm eq}_{N_k})\langle \Gamma_{N_k \rightarrow L_\alpha \eta}\rangle  - (Y_{\eta}-Y^{\rm eq}_{\eta})\langle \Gamma_{\eta \rightarrow L_\alpha N_1}\rangle \right. \nonumber \\
    &- \left. 
    2(Y^2_{\eta}-(Y^{\rm eq}_{\eta})^2)s\langle \sigma v\rangle_{\eta \eta \rightarrow {\rm SM SM}} - \sum^3_{m=1}(Y_{N_m}Y_{\eta}-Y^{\rm eq}_{N_m}Y^{\rm eq}_{\eta})s\langle \sigma v \rangle_{\eta N_m\rightarrow L{\rm SM}} \right], \nonumber \\
    \frac{dY_{N_1}}{dz} &= \frac{1}{zH(z)}\left[(Y_{\eta}-Y^{\rm eq}_{\eta})\langle \Gamma_{\eta \rightarrow L_\alpha N_1}\rangle - (Y_{N_1}Y_{\eta}-Y^{\rm eq}_{N_1}Y^{\rm eq}_{\eta})s\langle \sigma v \rangle_{\eta N_1\rightarrow L{\rm SM}} \right. \nonumber \\
    &- \left. \sum^3_{l=1}(Y_{N_1}Y_{N_l}-Y^{\rm eq}_{N_1}Y^{\rm eq}_{N_l})s\langle \sigma v \rangle_{N_l N_1\rightarrow {\rm SM SM}}\right].
\end{align}
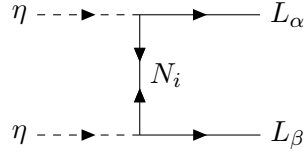
\begin{figure}
    \centering
\begin{tikzpicture}[/tikzfeynman/small]
    \begin{feynman}
    \vertex (i){$\eta$};
    \vertex [below = 1.6cm of i] (j){$\eta$};
    \vertex [right = 1.6cm of i] (v1);
    \vertex [right = 1.6cm of v1] (k){$L_\alpha$};
    \vertex [below = 1.6cm of v1] (v2);
    \vertex [below = 1.6cm of k] (m){$L_\beta$};
    \diagram*[small]{(i) -- [charged scalar] (v1),(v1) -- [fermion] (k),(v2) -- [ fermion] (m),(v2) -- [majorana,edge label'=$N_i$] (v1),(j) -- [charged scalar] (v2)};
    \end{feynman}
    \end{tikzpicture}    
    \caption{Feynman diagrams of the annihilation processes responsible for creating lepton asymmetry in fermion DM scenario.}
\label{fig:LL}
\end{figure}

Now, in this scenario along with the co-annihilation channels discussed earlier, the annihilation channels shown in Fig. \ref{fig:LL} also contribute to the asymmetry. The CP asymmetry coming from the interference of the tree (Fig. \ref{fig:LL}) and the loop (bottom two diagrams in Fig. \ref{fig:loop}) leading to $\mathcal{O}(y^6)$ are given as:
\begin{align}
    \epsilon_{\eta \eta} &= 8\sum_{ij} \left(\Im[(yy^\dagger)_{i1}(yy^\dagger)_{j1}(yy^\dagger)_{ij}]\right)\epsilon^{ann}_{ij} \nonumber \\
    &=  \sum_{j}\frac{m^4_1-m^4_j}{\Lambda^4_1\Lambda^2_j}\left(\Delta m^2_{1j}\sin(4\theta^R_{1j}) + (m^2_1 + m^2_j)\sinh(4\theta^I_{1j})\right)\epsilon^{ann}_{1j}, \nonumber \\
    \epsilon^{ann}_{1j} &= \frac{1}{16\pi}\left[1-r_1 - \frac{1}{2}(r_1-3)\ln\left[\frac{1+r_1}{3-r_1}\right]\right] \frac{\sqrt{r_jr_1}}{(1+r_j)(1+r_1)}\frac{1}{\mathcal{M}_{tree}}, \label{eq:anneps} \\
    \mathcal{M}_{tree} &= \sum_{ij} (yy^\dagger)^2_{ij}\frac{\sqrt{r_i}\sqrt{r_j}}{(1+r_i)1+r_j)}.
\end{align}
At this point one may notice that the asymmetry expression given in Eq. \eqref{eq:coanneps} is much more complicated than the one given in Eq. \eqref{eq:anneps} which is purely due to the presence of several coannihilating particles.. As pointed out earlier, in this scenario we would require at least one of the right handed neutrinos to be lighter than the scalar doublet $\eta$ whose annihilations are responsible for creating the asymmetry. In our case we have considered only $N_1$ to be lighter than $\eta$ and the rest to be heavier. Alternatively, if $N_{2,3}$ are lighter than $\eta$, then their (co)annihilations can contribute more to the generation to the asymmetry. For our chosen benchmark points, as shown in table \ref{tab:BPF}, the contribution of $N_{2,3}$ annihilations to lepton asymmetry is sub-dominant compared to $\eta$ annihilations as well as $\eta-N_k\; (k=2,3)$ coannihilations. In fact in the fermion DM scenario, both $\eta-N_k\; (k=2,3)$ coannihilations (shown in Fig. \ref{fig:coasym}) as well as $\eta$ annihilations (shown in Fig. \ref{fig:LL}) can contribute to lepton asymmetry. It is worthwhile to note that  the $\eta$ annihilations shown in Fig. \ref{fig:LL} can not contribute to lepton asymmetry in the scalar DM scenario due to the absence of "on-shell" condition for loop particles.

The washout effects in this scenario are categorised as follows :
\begin{itemize}
\item $\bf \Delta L = 2 $: there are two processes of this type:
\begin{enumerate}
\item $\eta \eta \rightarrow L L$ and $L \eta \rightarrow \overline{L} \eta$ where the former is also responsible for the source of asymmetry while the latter is purely wash-out.
\end{enumerate}
\item $\bf \Delta L = 1 $: there are two main sources of such washout processes:
\begin{enumerate}
\item the inverse decay of $\Gamma(N_k \rightarrow L \eta)$ and $\Gamma(\eta \rightarrow L N_k)$, and the second one is purely wash-out  not contributing to the asymmetry.
\item  the inverse process of co-annihilation $N_k \eta \rightarrow L,X(=\gamma,W,h)$
\end{enumerate}
\end{itemize}

\begin{table}[]
    \centering
    \begin{tabular}{|c|c|}
    \hline 
    Parameter & Required parameter for Correct Relic \\
     \hline   
      $\mu_\eta$  &  870 GeV\\
     $M_{\eta_R} = M_{\eta_I}$ & 870 GeV\\
      $M_{\eta_\pm} $ & 876.3 GeV\\
        $M_{N_1}$ &  $M_{\eta_\pm} - \Delta m$ GeV \\
        $\Delta m$ &  10 GeV\\
        $M_{N_2}$ & 1 TeV\\
        $M_{N_3}$ & 2 TeV\\
        $\lambda_1$ & 0.253\\
        $\lambda_3$ & 1.48\\
        $\lambda_4$ & -1.48 \\
        $\lambda_5$ & $1\times10^{-5}$ \\
        $\lambda_2$ & 1 \\
       $\theta^R_{ij}$ & $-\frac{\pi}{4}\omega$ \\
        $\theta^I_{12}$ & $\frac{3\pi}{4}\omega$\\
        $\theta^I_{13}=\theta^I_{23}$ & $\frac{\pi}{4}\omega$\\
        \hline
    \end{tabular}
    \caption{The numerical values of the parameters chosen for generating correct Fermion DM relic and baryon asymmetry. We denote it as benchmark point 2 (BP2). The fraction $\omega$ is to ensure to get the correct baryonic asymmetry. For Normal (Inverted) Hierarchy $\omega = 0.85(0.45)$. }
    \label{tab:BPF}
\end{table}

\begin{figure}[!ht]
    \centering
    
    \begin{tabular}{lr}
    \includegraphics[width=0.5\textwidth]{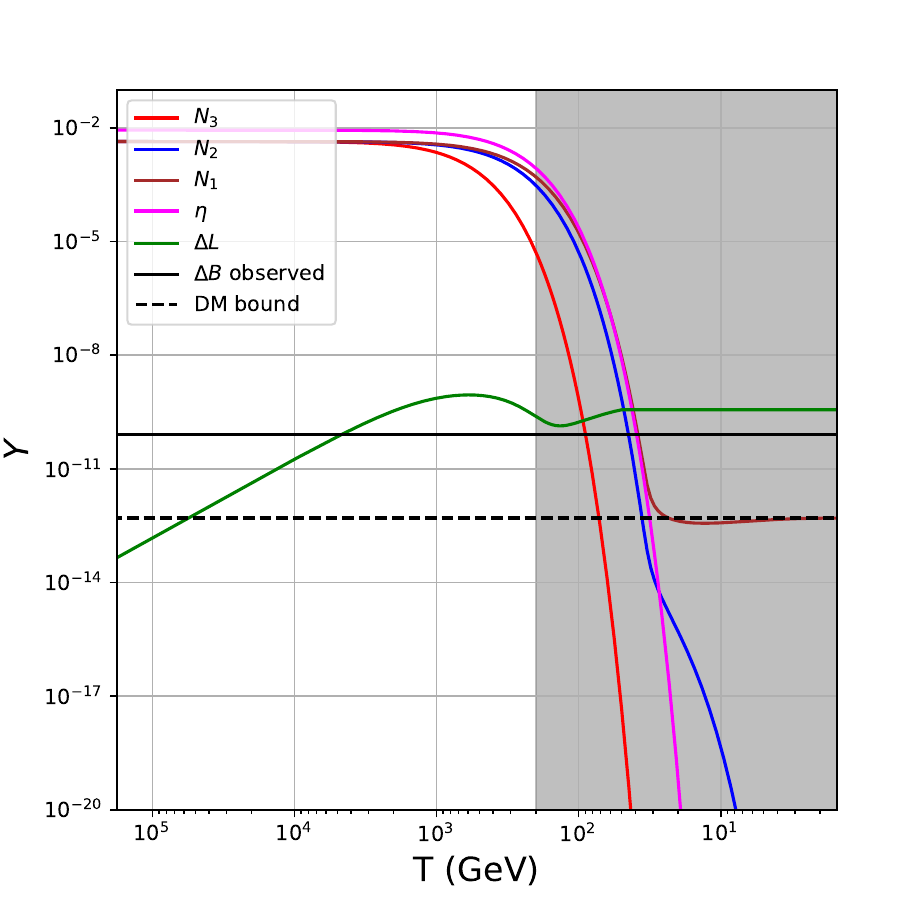}&
     \includegraphics[width=0.5\textwidth]{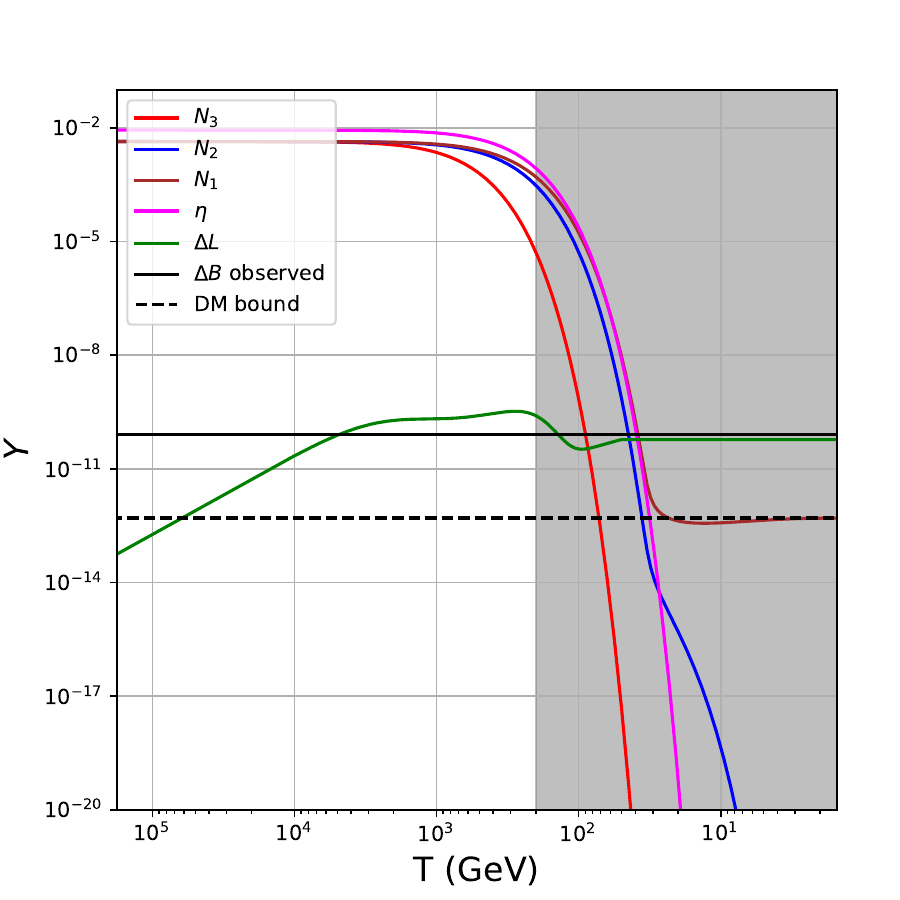}\\
     \includegraphics[width=0.5\textwidth]{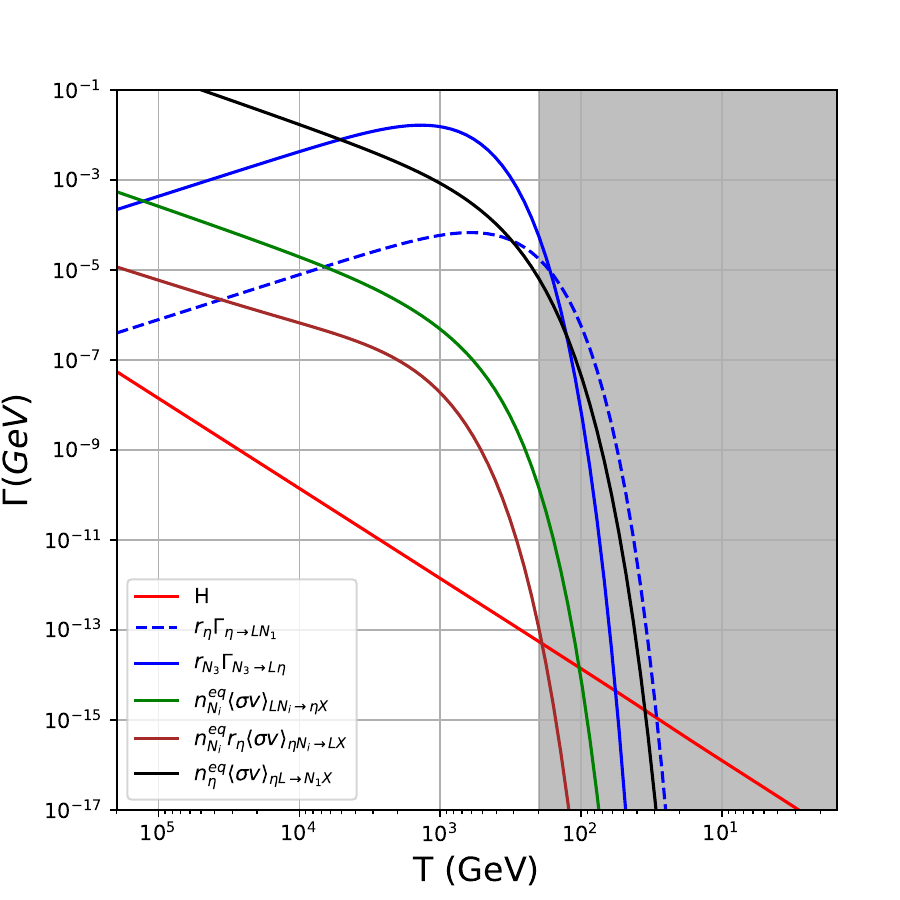}&
    \includegraphics[width=0.5\textwidth]{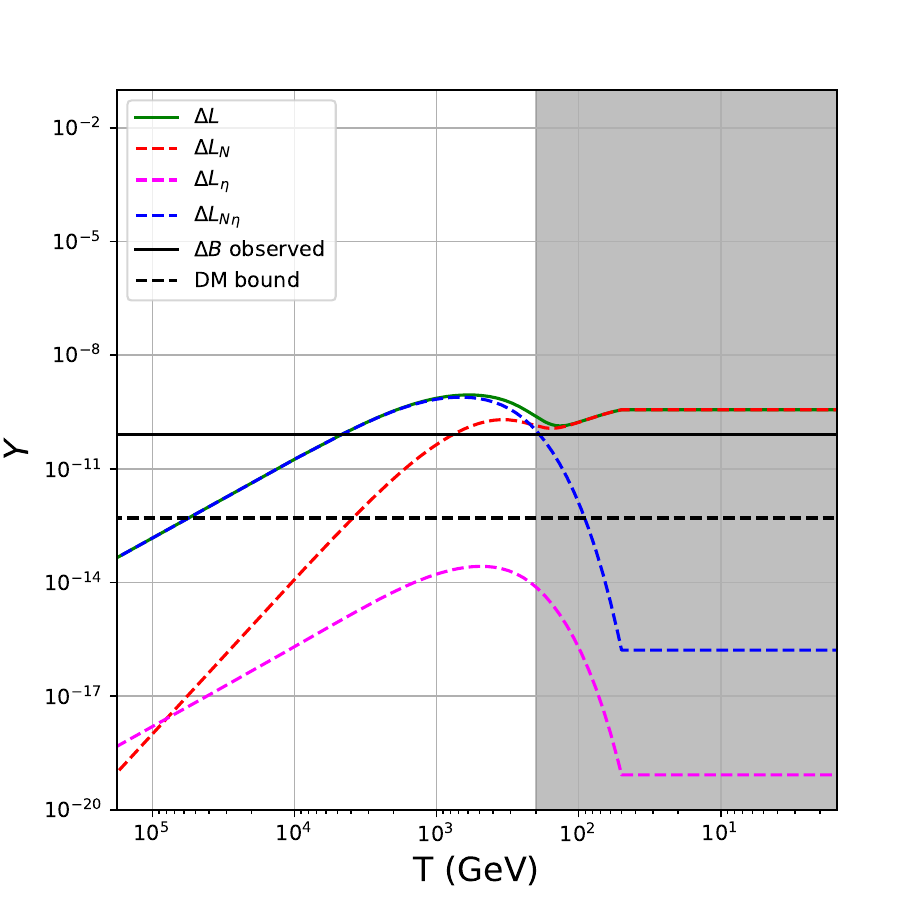}
    \end{tabular}
    \caption{Top panel: comoving number densities  of $Z_2$ odd particles and lepton asymmetry as a function of temperature for fermion DM scenario. The left (right) panel corresponds to the normal (inverted) hierarchy of neutrino mass spectrum.Bottom panel: rates of individual wash-out processes along with the Hubble rate (left panel); individual (co)annihilation contributions of each channel to the asymmetry. The solid (dashed) black line corresponds to the baryon asymmetry (DM abundance) observed at present epoch ($T \sim 0$). Shaded regions represent the epochs after sphaleron freeze out.}
    \label{fig:LLF}
\end{figure}

In Fig. \ref{fig:LLF}, we plot the predictions for relic abundance of all $Z_2$ odd particles, taking part in generating the lepton asymmetry along with the generated asymmetry $\Delta L$, as functions of temperature, for fermion DM scenario. The upper left panel corresponds to the NH of neutrino masses, whereas the upper right panel to the IH of neutrino masses. In the lower panel, we compare the relative contributions to the lepton asymmetry: from $\eta-\eta$ annihilations $(\Delta L_{\eta})$, from $\eta-N_k$ coannihilations $(\Delta L_{N \eta})$, from $N_2$ decay $(\Delta L_N)$, as functions of temperature.
Similar to the scenario of scalar DM, here also the lepton asymmetry grows as temperature cools due to contributions from the annihilations and coannihilations and saturate around the temperature where the processes responsible for creating the asymmetry tend to go out of equilibrium. Also, the lightest right handed neutrino freezes out to give the required relic abundance for DM in the universe. 
The shaded regions in the panels correspond to the temperature below
which the sphaleron processes become inoperative.

The first bump in the curve for the lepton asymmetry shown in upper panel plots of Fig. \ref{fig:LLF} arises from the coannihilation diagrams 
while the decay contribution enters later and further increases the asymmetry at later stages. Here we notice that the 
mass of the lightest right handed fermion is very close to the $Z_2$ odd scalar 
counterparts in order to enhance the coannihilations. This is evident from the bottom panel plot of Fig. \ref{fig:LLF} which shows the co-annihilation among the lightest 
$N$ and $Z_2$ odd scalar components contribute dominantly to lepton asymmetry.  Along with that we have an interesting feature of $\lambda_5$. In this we have seen that for particular parameter set as shown in table \ref{tab:lambda5F}, the upper bound on $\lambda_5$ is set by the requirement of the required leptonic asymmetry needed to explain the observed baryonic asymmetry in the Universe and the lower bound is set by the Lepton flavor violating process ${\rm BR}(\mu \rightarrow e \gamma) < 4.2 \times 10^{-13}$ \cite{TheMEG:2016wtm}, as we discuss below. Since the dark matter candidate is a fermion singlet in this case, the parameter $\lambda_5$ is not constrained by dark matter direct detection.
\begin{table}
    \centering
    \begin{tabular}{|l|c|}
    \hline
        & $\lambda_5$ \\ \hline
    NH    &  $(0.027-8)\times 10^{-5}$\\ \hline
    IH    &  $(0.029-6.7)\times 10^{-5}$\\ \hline
    \end{tabular}
    \caption{The range of $\lambda_5$ allowed by phenomenological requirements of satisfying charged lepton flavour violation bounds and generating the required lepton asymmetry.}
    \label{tab:lambda5F}
\end{table} 
While there exist some wash-out effects for the coannihilating processes, there 
is not much of that for the annihilation processes, as the processes are already out 
of equilibrium when wash-out becomes effective. In this case, large Yukawa 
couplings are required to achieve successful leptogenesis, which in turn leads
to very small $\lambda_5$, but yet does not affect fermion DM phenomenology 
much. The Yukawa couplings can be represented by the following matrix for the 
chosen benchmark point in this scenario.
\begin{align}
y\vert_{NH} &= \begin{pmatrix}
1.01 - 1.14i &	2.43 + 1.09i &	1.44 + 1.8i \\	
-0.15 - 1.21i &	1.92 - 9.29i &	1.67 + 0.13i \\	
-1.40 - 0.61i &	2.55 - 2.6i &	1.37 - 1.81i \\
\end{pmatrix} \times 10^{-2}, \\
y\vert_{IH} &= \begin{pmatrix}
1.88 - 3.05i &	1.28 + 0.55i &	-1.25 - 0.45i \\	
-1.12 - 2.92i &	1.04 - 0.9i &	-1.01 + 0.92i \\	
-3.87 - 0.79i &	-0.59 - 1.53i &	0.61 + 1.49i \\	
\end{pmatrix} \times 10^{-2}.
\end{align}

%

\begin{figure}
\centering
    \includegraphics[width=0.48\textwidth]{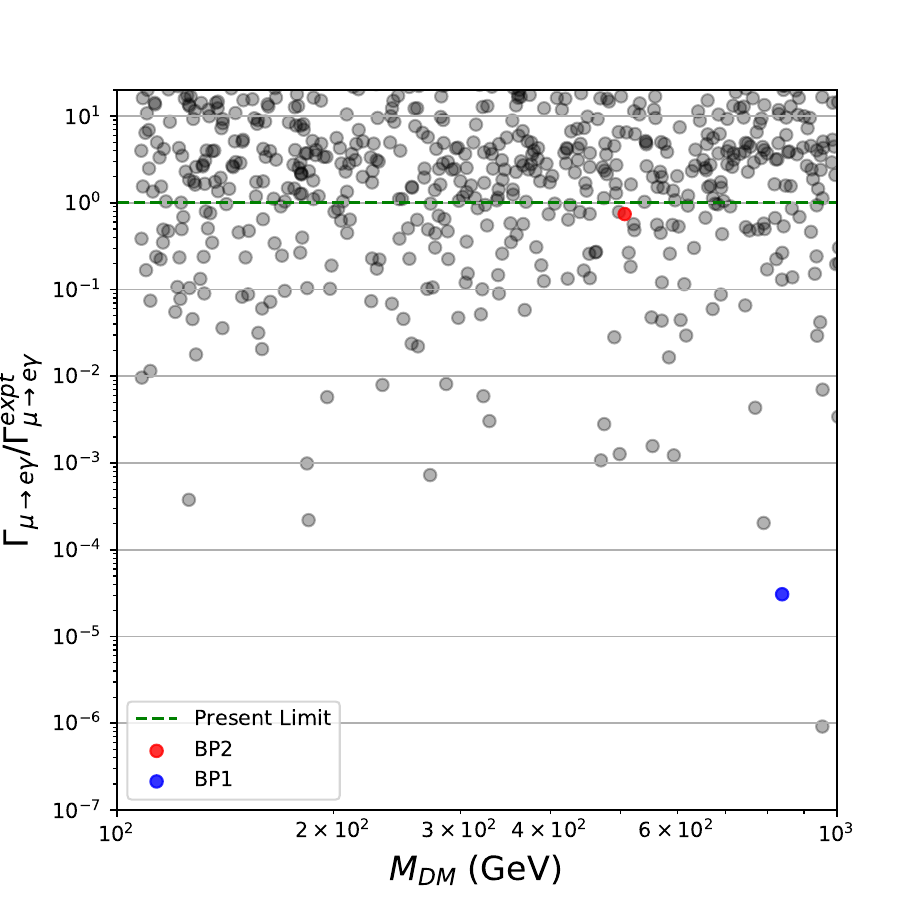} 
    \includegraphics[width=0.48\textwidth]{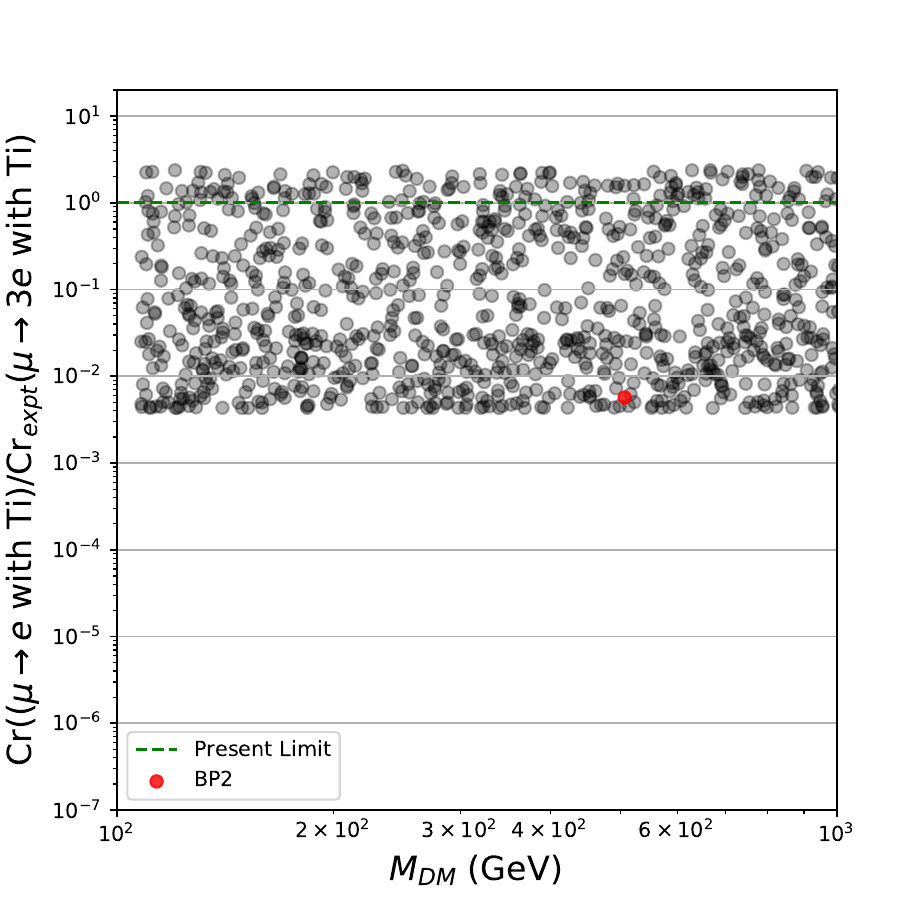} \\
     \includegraphics[width=0.48\textwidth]{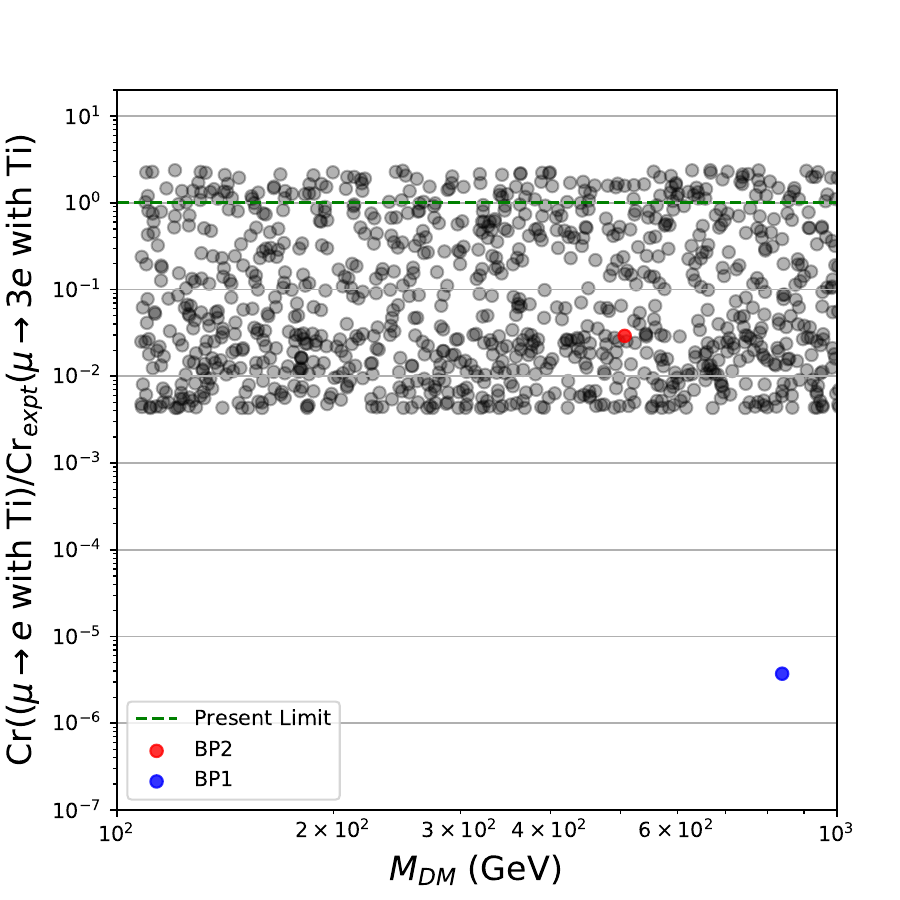} 

 \caption{Predictions for LFV processes for $10^2~{\rm GeV} < M_{\rm DM}< 10^3~{\rm GeV}$. The two benchmark points are highlighted with red and blue coloured points. The $\mu_\eta$ are varied from 100 GeV to 1 TeV, and the other parameters are taken as presented in table \ref{tab:BPF}.}
 \label{lfvplot}
\end{figure}

We then use the \texttt{SPheno 3.1} interface to check the constraints from flavour data. We particularly focus on three charged lepton flavour violating (LFV) decays namely, $\mu \rightarrow e \gamma, \mu \rightarrow 3e$ and $\mu \rightarrow e$ (Ti) conversion that have strong current limit as well as good future sensitivity \cite{Toma:2013zsa}. The present bounds are: ${\rm BR}(\mu \rightarrow e \gamma) < 4.2 \times 10^{-13}$ \cite{TheMEG:2016wtm},  ${\rm BR}(\mu \rightarrow 3e) < 1.0 \times 10^{-12}$ \cite{Bellgardt:1987du}, ${\rm CR} (\mu, \rm Ti \rightarrow e, \rm Ti) < 4.3 \times 10^{-12}$ \cite{Dohmen:1993mp}. While the future sensitivity of the first two processes are around one order of magnitude lower than the present branching ratios, the $\mu$ to $e$ conversion (Ti) sensitivity is supposed to increase by six order of magnitudes \cite{Toma:2013zsa} making it a highly promising test to confirm or rule out different TeV scale BSM scenarios. It should be noted that such charged LFV process arises in the SM at one loop level and remains suppressed by the smallness of neutrino masses, much beyond the current and near future experimental sensitivities. Therefore, any experimental observation of such processes is definitely a sign of BSM physics, like the one we are studying here. We show the predictions for LFV processes in our model in Fig. \ref{lfvplot}, also highlighting the second benchmark point (BP2) mentioned above. The similar contributions for the BP1 scenario remain far more suppressed due to smallness of the corresponding Yukawa couplings. The scatter plot in Fig. \ref{lfvplot} is obtained by only varying $\mu_\eta$ from 100 GeV to 1 TeV, and fixing the other parameters with the values presented in table \ref{tab:BPF}. It can be seen that some part of the parameter space, specially the region which generates correct DM abundance, lies close to the current experimental limits. As mentioned earlier, this bound on ${\rm BR}(\mu \rightarrow e \gamma)$ decides the lower bound on the parameter $\lambda_5$ shown in table \ref{tab:lambda5F}. If $\lambda_5$ is lower than the one chosen in BP2, the corresponding Yukawa couplings will be bigger (from Casas-Ibarra parametrisation) enhancing the decay rate. For $\mu \rightarrow e \gamma$, the latest MEG 2016 limit \cite{TheMEG:2016wtm} can already rule out several points. The promising future sensitivity of the $\mu$ to $e$ conversion (Ti) will be able to explore most part of the parameter space.

\section{Conclusion}
\label{sec:level5}
We have proposed a scenario where baryogenesis via leptogenesis can be achieved through annihilations and coannihilations of particles belonging to a $Z_2$ odd sector including t-channel processes. We have considered a popular model known as the scotogenic model to implement the idea, and addressed the the possibility of explaining the coincidence of DM abundance and baryon asymmetry in the present universe along with non-zero neutrino masses. Pointing out two different possible scenarios corresponding to scalar and fermion DM respectively, we show the non-trivial role played by t-channel annihilation as well as coannihilation processes between different $Z_2$ odd particles. For the two benchmark points chosen in our work, we could obtain successful leptogenesis along with other requirements like DM relic, DM direct detection and light neutrino masses for $M_{\rm DM} \sim 870$ GeV whereas vanilla leptogenesis in scotogenic model works for $M_1 \geq 10$ TeV. Another interesting feature is the testability of the model at DM direct detection and rare decay experiments. Even though the particle spectrum is in a few $\mathcal{O}(100)$ GeV regime or above, away from the reach of current collider experiments, the model can still be tested at near future run of rare decay experiments looking for charged lepton flavour violation like $\mu \rightarrow e \gamma, \mu \rightarrow 3 e$, $\mu$ to $e$ conversion etc. We highlight our interesting results by adopting benchmark points here and leave a detailed numerical analysis of this scenario to an upcoming work.

\appendix

\section{Details of the Boltzmann Equations}
\label{sec:appen1}
In this section, we  present the Boltzmann Equation (BE) used in our analysis, in detail. The standard BE is written as follows:
\begin{align}
\frac{d X_\alpha}{dz} &= \frac{\Delta n_\alpha}{sH(z)z} \frac{X_{i_1}X_{i_2}\cdots}{X^{\rm eq}_{i_1}X^{\rm eq}_{i_1}\cdots}\gamma^{\rm eq}(i_1 + i_2 \cdots \rightarrow f_1 + f_2 + \cdots),
\label{eq:BE1}
\end{align}
where $X_{\alpha}=n_{\alpha}/s$ is the comoving number density, $z = M_r/T$ ($M_r$ is the reference scale which is same as the mass of the decaying particle in usual leptogenesis/baryogenesis while being equal to the mass of dark matter in the case of dark matter freeze-out ) and $\Delta n_\alpha$ is the change of number of particle in the process considered. Along with that, the Hubble expansion rate, $H(z)$, and entropy density, $s$, are defined respectively as 
\begin{align}
H(z) &= \sqrt{\frac{4\pi^3g_*}{45}}\frac{T^2}{M_{\rm PL}}  = \sqrt{\frac{4\pi^3g_*}{45} }\frac{M_r^2}{M_{\rm PL}z^2} = \frac{H(M_r)}{z^2}, \\
s &= g_* \frac{2\pi^2}{45}T^3.
\end{align}
Now, the collision term $\gamma^{\rm eq}(i_1 + i_2 \cdots \rightarrow f_1 + f_2 + \cdots)$ are of two types: 1) decay  and 2) annihilation processes. For decay
process it is given as:
\begin{align}
\gamma^{\rm eq}_D(i \rightarrow f_1 + f_2 + \cdots) = \frac{g_{i} M^2_{i}}{2\pi^2}TK_1(M_{i}/T)\Gamma(i \rightarrow f_1 + f_2 + \cdots),
\end{align}
and for the scattering processes it is given as
\begin{align}
\gamma^{\rm eq}_{s}(i_1 + i_2 \rightarrow f_1 + f_2 + \cdots) &= \frac{g_{i_1}g_{i_2}T}{64\pi^4}\int^\infty_{s_{min}} ds \frac{2\lambda(s,m^2_{i_1},m^2_{i_2})}{s}\sigma \sqrt{s}K_1(\sqrt{s}/T),
\end{align}
where $K_n$ is the $n$th order Modified Bessel function of second kind, $\lambda(a,b,c) = a^2 + b^2 + c^2 -2ab - 2ac - 2bc$ and $s_{min} $ = Max$\{(m_{i_1} + m_{i_2})^2,(m_{f_1} + m_{f_2} + \cdots)^2\}$. 

\begin{figure}
\centering
\begin{tabular}{lr}
     \begin{tikzpicture}[/tikzfeynman/small]
     \begin{feynman}
     \vertex (i){$\eta$};
     \vertex [below = 2.cm of i] (j){$L_\alpha$};
     \vertex [below right= 1.414cm of i] (v1);
     \vertex [right = 1.cm of v1] (v2);
     \vertex [right = 3.cm of i] (m){$\eta$};
     \vertex [below = 2.cm of m] (o){$L_\alpha$};
     \diagram*[small]{(i) -- [anti charged scalar] (v1),(v1) -- [anti majorana,edge label = $N_i$] (v2),(v2) -- [charged scalar] (m),(v2) -- [anti fermion] (o),(j) -- [fermion] (v1)};
     \end{feynman}
     \end{tikzpicture}
     &
     \begin{tikzpicture}[/tikzfeynman/small]
     \begin{feynman}
     \vertex (i){$\eta$};
     \vertex [right = 1.cm of i](v1);
     \vertex [above right = 1.cm of v1] (j){$L_\alpha$};
     \vertex [below right = 1.cm of v1] (k){$N_i$};
     \diagram*[small]{(i) -- [charged scalar](v1),
     (v1) -- [fermion] (j),(v1) -- [fermion] (k)};
     \end{feynman}
     \end{tikzpicture}
\end{tabular}
\caption{Diagrams leading to  $\gamma^{\rm eq}_{s}(\eta L \rightarrow \eta \overline{L})$(left) and $\gamma_D(\eta \rightarrow N_1 L)$(right), respectively.}
 \label{fig:beproc}
\end{figure}
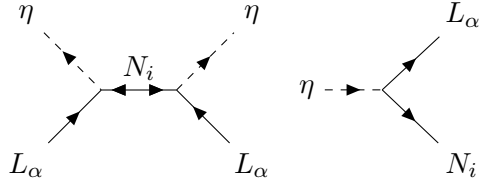

Now, we present the explicit expressions of the BEs for $N_k$ and $\eta$ in the case of scalar dark matter. 
\begin{align}
    \frac{dY_{N_k}}{dz} &= -\frac{1}{szH(z)}\left[\left(\frac{Y_{N_k}}{Y^{\rm eq}_{N_k}}-1\right)\gamma_D(N_k \rightarrow L_\alpha \eta) + \left(\frac{Y_{N_k}}{Y^{\rm eq}_{N_k}}\frac{Y_{\eta}}{Y^{\rm eq}_{\eta}}-1\right)\gamma^{\rm eq}_{s}(\eta N_k\rightarrow L{\rm SM}) \right. \nonumber \\
    &+ \left. \sum_{l=1}^{3}\left(\frac{Y_{N_k}Y_{N_l}}{Y^{\rm eq}_{N_k}Y^{\rm eq}_{N_l}}-1\right)\gamma^{\rm eq}_{s}(N_l N_k\rightarrow {\rm SM SM})\right], \nonumber \\
    \frac{dY_{\eta}}{dz} &= \frac{1}{szH(z)}\left[ 
    \left(\frac{Y_{N_k}}{Y^{\rm eq}_{N_k}}-1\right)\gamma_D(N_i \rightarrow L_\alpha \eta) 
     - 2\left(\frac{Y^2_{\eta}}{(Y^{\rm eq}_{\eta})^2}-1\right)\gamma^{\rm eq}_{s}(\eta \eta \rightarrow {\rm SM SM})\right. \nonumber \\
    &- \left. \sum^3_{m=1}\left(\frac{Y_{N_m}Y_{\eta}}{Y^{\rm eq}_{N_m}Y^{\rm eq}_{\eta}}-1 \right)\gamma^{\rm eq}_{s}(\eta N_m\rightarrow L{\rm SM}) \right].
\end{align}
And similarly expressions of BEs for $N_k$ and $\eta$  in the case of  fermion dark matter.
\begin{align}
  \frac{dY_{N_k}}{dz} &= -\frac{1}{szH(z)}\left[\left(\frac{Y_{N_k}}{Y^{\rm eq}_{N_k}}-1\right)\gamma_D(N_k \rightarrow L_\alpha \eta) + \left(\frac{Y_{N_k}}{Y^{\rm eq}_{N_k}}\frac{Y_{\eta}}{Y^{\rm eq}_{\eta}}-1\right)\gamma^{\rm eq}_{s}(\eta N_k\rightarrow L{\rm SM}) \right. \nonumber \\
    &+ \left. \sum_{l=1}^{3}\left(\frac{Y_{N_k}Y_{N_l}}{Y^{\rm eq}_{N_k}Y^{\rm eq}_{N_l}}-1\right)\gamma^{\rm eq}_{s}(N_l N_k\rightarrow {\rm SM SM})\right], \quad \textrm{ for} ~k = 2,3 \nonumber \\
    \frac{dY_{N_1}}{dz} &= \frac{1}{szH(z)}\left[\left(\frac{Y_{\eta}}{Y^{\rm eq}_{\eta}}-1\right)\gamma_D(\eta \rightarrow L_\alpha N_1) - \left(\frac{Y_{N_1}}{Y^{\rm eq}_{N_1}}\frac{Y_{\eta}}{Y^{\rm eq}_{\eta}}-1\right)\gamma^{\rm eq}_{s}(\eta N_1\rightarrow L{\rm SM}) \right. \nonumber \\
    &- \left. \sum^3_{l=1}\left(\frac{Y_{N_1}Y_{N_l}}{Y^{\rm eq}_{N_1}Y^{\rm eq}_{N_l}}-1\right)\gamma^{\rm eq}_{s}(N_l N_1\rightarrow {\rm  SM SM})\right], \nonumber \\
    \frac{dY_{\eta}}{dz} &= \frac{1}{szH(z)}\left[ 
    \sum^3_{k=2}\left(\frac{Y_{N_k}}{Y^{\rm eq}_{N_k}}-1\right)\gamma_D(N_k \rightarrow L_\alpha \eta)  - \left(\frac{Y_{\eta}}{Y^{\rm eq}_{\eta}}-1\right)\gamma_D(\eta \rightarrow L_\alpha N_1) \right. \nonumber \\
    &- \left. 
    2 \left(\frac{Y^2_{\eta}}{(Y^{\rm eq}_{\eta})^2}-1\right)\gamma^{\rm eq}_{s}(\eta \eta \rightarrow {\rm SM SM}) - \sum^3_{m=1} \left(\frac{Y_{N_m}}{Y^{\rm eq}_{N_m}}\frac{Y_{\eta}}{Y^{\rm eq}_{\eta}}-1\right)\gamma^{\rm eq}_{s}(\eta N_m\rightarrow L{\rm SM}) \right].
\end{align}
Finally the expressions of BEs for the lepton numbers as follows:
\begin{align}
 \frac{dY_L}{dz} &=  \frac{1}{sH(z)z}\left[\sum_j\left( \left(\frac{Y_{N_j}}{Y^{eq}_{N_j}} - \frac{Y_L}{Y^{\rm eq}_L}\right)(1+\epsilon_{N_j})\gamma_D(N_j \rightarrow L_\alpha \eta)\right) \right. \nonumber \\
     &+  2\left(\frac{Y^2_{\eta}}{(Y_{\eta}^{\rm eq})^2} - \frac{Y^2_L}{(Y_L^{\rm eq})^2}\right)(1 + \epsilon_{\eta \eta})\gamma^{\rm eq}_{s}(\eta \eta \rightarrow L L) \nonumber \\
     &+  \sum_i \left(\frac{Y_{\eta}Y_{N_i}}{Y^{\rm eq}_{\eta}Y^{\rm eq}_{N_i}} - \frac{Y_L}{Y^{\rm eq}_L}\right)(1 + \epsilon_{N_i \eta} )\gamma^{\rm eq}_{s}(\eta N_i\rightarrow L{\rm SM})  \nonumber \\
     &+ \left. \sum_i\left(\frac{Y^{eq}_{Ni}}{Y^{eq}_{N_i}}\frac{Y^{eq}_{X}}{Y^{eq}_{X}}-\frac{Y^{eq}_{\eta}}{Y^{eq}_{\eta}}\frac{Y_L}{Y^{eq}_L}\right)(1+\epsilon_{\eta L})\gamma^{eq}_s(N_iX\rightarrow \eta L)\right. \nonumber \\
     &+ \left. \sum_i\left(\frac{Y^{eq}_{\eta}}{Y^{eq}_{\eta}}\frac{Y^{eq}_{X}}{Y^{eq}_{X}}-\frac{Y^{eq}_{Ni}}{Y^{eq}_{N_i}}\frac{Y_L}{Y^{eq}_L}\right)(1+\epsilon_{N_i L})\gamma^{eq}_s(\eta X\rightarrow N_i L)\right. \nonumber \\
     &\left.-\frac{Y_{\Delta L}}{Y^{\rm eq}_L}\gamma^{\rm eq}_{s}(\eta L \rightarrow \eta \overline{L}) + \left(\frac{Y_\eta}{Y^{\rm eq}_\eta} - \frac{Y_L}{Y^{\rm eq}_L}\right)\gamma_D(\eta \rightarrow N_1 L)\right], \nonumber \\
   & \label{eq:Lep}
   \end{align}
   \begin{align}
    \frac{dY_{\bar{L}}}{dz} &=  \frac{1}{sH(z)z}\left[\sum_j\left( \left(\frac{Y_{N_j}}{Y^{\rm eq}_{N_j}} - \frac{Y_{\bar L}}{Y^{\rm eq}_L}\right)(1-\epsilon_{N_j})\gamma_D(N_j \rightarrow L_\alpha \eta)\right) \right. \nonumber \\
     &+  2\left(\frac{Y^2_{\eta}}{(Y_{\eta}^{\rm eq})^2} - \frac{Y^2_{\bar L}}{(Y_L^{\rm eq})^2}\right)(1 - \epsilon_{\eta \eta})\gamma^{\rm eq}_{s}(\eta \eta \rightarrow L L) \nonumber \\
     &+  \sum_i \left(\frac{Y_{\eta}Y_{N_i}}{Y^{\rm eq}_{\eta}Y^{\rm eq}_{N_i}} - \frac{Y_{\bar L}}{Y^{\rm eq}_L}\right)(1 - \epsilon_{N_i \eta} )\gamma^{\rm eq}_{s}(\eta N_i\rightarrow L{\rm SM})  \nonumber \\
     &+ \left. \sum_i\left(\frac{Y^{eq}_{Ni}}{Y^{eq}_{N_i}}\frac{Y^{eq}_{X}}{Y^{eq}_{X}}-\frac{Y^{eq}_{\eta}}{Y^{eq}_{\eta}}\frac{Y_{\bar L}}{Y^{eq}_L}\right)(1-\epsilon_{\eta L})\gamma^{eq}_s(N_iX\rightarrow \eta L)\right. \nonumber \\
     &+ \left. \sum_i\left(\frac{Y^{eq}_{\eta}}{Y^{eq}_{\eta}}\frac{Y^{eq}_{X}}{Y^{eq}_{X}}-\frac{Y^{eq}_{Ni}}{Y^{eq}_{N_i}}\frac{Y_{\bar L}}{Y^{eq}_L}\right)(1-\epsilon_{N_i L})\gamma^{eq}_s(\eta X\rightarrow N_i L)\right. \nonumber \\
     &\left.+\frac{Y_{\Delta L}}{Y^{\rm eq}_L}\gamma^{\rm eq}_{s}(\eta L \rightarrow \eta \overline{L}) + \left(\frac{Y_\eta}{Y^{\rm eq}_\eta} - \frac{Y_{\bar L}}{Y^{\rm eq}_L}\right)\gamma_D(\eta \rightarrow N_1 L)\right].  \label{eq:antiLep}
\end{align}
where 
the leptonic comoving number density are defined as $Y_L = Y^{\rm eq}_L+ Y_{\Delta L}/2, Y_{\overline{L}} = Y^{\rm eq}_L - Y_{\Delta L}/2$. 
The rates $\gamma_D(N_j\rightarrow L_\alpha \eta)$ are the decay process considered in \cite{Hugle:2018qbw} and $\gamma^{\rm eq}_s(\eta \eta \rightarrow LL)$ are shown in Fig. \ref{fig:LL}, whereas $\gamma^{\rm eq}_{s}(\eta N_i\rightarrow L{\rm SM})$ and $\gamma^{\rm eq}_s(N_iX\rightarrow \eta L)$ are shown in Fig. \ref{fig:coasym}. The processes $\gamma^{\rm eq}_s(N_iX\rightarrow \eta L)$ and $\gamma^{\rm eq}_s(\eta X\rightarrow N_i L)$ are the same as Fig. \ref{fig:coasym} by just interchanging the one of the initial with the final particle. And finally the processes $\gamma^{\rm eq}_{s}(\eta L \rightarrow \eta \overline{L})$ and $\gamma_D(\eta \rightarrow N_1 L)$ are shown in Fig. \ref{fig:beproc}. There are no other further processes which contributes in the above Boltzmann equation.
Now taking the difference between Eq. \eqref{eq:Lep} and Eq. \eqref{eq:antiLep} and keeping the terms asymmetry to leading order i.e $Y^2_L \simeq (Y^{\rm eq}_L)^2 + Y^{\rm eq}_LY_{\Delta L}$, one would get the final Boltzmann equation for asymmetry eq.\eqref{eq:asym}
\begin{align}
    \frac{dY_{\Delta L}}{dz} &= \frac{1}{sH(z)z}\left[\sum_j\left( \left(\frac{Y_{N_j}}{Y^{eq}_{N_j}} - 1\right)\epsilon_{N_j}\gamma_D(N_j \rightarrow L_\alpha \eta) -  \frac{Y_{\Delta L}}{Y^{\rm eq}_L}\gamma_D(N_j \rightarrow L_\alpha \eta) \right)\right. \nonumber \\
    &+  2\left(\frac{Y^2_{\eta}}{(Y_{\eta}^{\rm eq})^2} - 1\right)\epsilon_{\eta \eta}\gamma^{\rm eq}_{s}(\eta \eta \rightarrow L L) -  2\frac{Y_{\Delta L}}{Y^{\rm eq}_L}\gamma^{\rm eq}_{s}(\eta \eta \rightarrow L L) \nonumber \\
    &+  \sum_i \left(\left(\frac{Y_{\eta}Y_{N_i}}{Y^{\rm eq}_{\eta}Y^{\rm eq}_{N_i}} - 1\right)\epsilon_{N_i \eta}\gamma^{\rm eq}_{s}(\eta N_i\rightarrow L{\rm SM})  - \frac{Y_{\Delta L}}{Y^{\rm eq}_L}\gamma^{\rm eq}_{s}(\eta N_i\rightarrow L{\rm SM})\right)\nonumber \\
    &- \sum_i\frac{Y_{\Delta L}}{Y^{\rm eq}_L}\left(\gamma^{eq}_s(N_iX\rightarrow \eta L) + \gamma^{eq}_s(\eta X\rightarrow N_i L) \right)\nonumber \\
     &\left.-2\frac{Y_{\Delta L}}{Y^{\rm eq}_L}\gamma^{\rm eq}_{s}(\eta L \rightarrow \eta \overline{L}) - \frac{Y_{\Delta L}}{Y^{\rm eq}_L}\gamma_D(\eta \rightarrow N_1 L)\right], \label{eq:BE}
\end{align}
Now, one would notice that in Eqs. \eqref{eq:Lep},\eqref{eq:antiLep} and \eqref{eq:BE} the index $j$ runs from $1-3$ if scalar is the Dark Matter in which case the last decay term $\gamma_D(\eta \rightarrow N_1 L)=0$. But, for the case of $N_1$ as the dark matter $j$ runs from $2-3$ and the last decay term $\gamma_D(\eta \rightarrow N_1 L)\neq0$. Now, one may notice that from CPT invariance and unitarity terms proportional to  $\epsilon_{N_i L}$ and $\epsilon_{\eta L}$ cancel out exactly. 
Hence, the effects coming from the $\gamma^{\rm eq}_s(N_iX\rightarrow \eta L)$ and $\gamma^{\rm eq}_s(\eta X\rightarrow N_i L)$ 
only contribute to the wash-out but it is suppressed compared to the inverse decay processes as shown in Fig. \ref{fig:LLS} and Fig. \ref{fig:LLF}.

\section{Details of the asymmetry}
\label{sec:appen3}

In this section we give the details of the asymmetry shown in Eq. 
\eqref{eq:coanneps}. We would first start with the the basic general expression 
giving the asymmetry shown as follows:
\begin{align}
    \delta = 4\Im[\mathcal{C}^{*}_0\mathcal{C}_1]\Im[\mathcal{A}_0\mathcal{A}_1]|\mathcal{W}|^2,
\end{align}
where $\mathcal{W}$ corresponds to the wavefunction of the incoming and outgoing particles, $\mathcal{C}$'s corresponds to the couplings of the tree ($\mathcal{C}_0$) and loop ($\mathcal{C}_1$) and $\mathcal{A}$'s corresponds the rest of the amplitude respectively. 

Now, starting with the tree level amplitudes 
\begin{align}
    \mathcal{M}_{0} &= i\frac{x^\dagger_\alpha\bar{\sigma}(p_1+p_2)\sigma y^{\dagger}_{N_i}}{s} + i\frac{x^\dagger_\alpha y^{\dagger}_{N_i}(2p_1-p_3)^\mu}{t - m^2_\eta},
\end{align}
where the amplitude $\mathcal{M}_{0}$ corresponds to the the tree diagram shown in Fig. \ref{fig:coasym}. The corresponding amplitudes for the loop correction are given as follows 
\begin{align}
    \mathcal{M}_{1} &= \frac{x^\dagger_\alpha \bar{\sigma}^\mu(p_1 + p_2)\sigma p_1\bar{\sigma}x_{N_i}m_{N_j}}{s}\mathcal{C}_1 \nonumber + \frac{x^\dagger_\alpha \bar{\sigma}^\mu(p_1 + p_2)\sigma p_2\bar{\sigma}x_{N_i}m_{N_j}}{s}\mathcal{C}_2 \nonumber \\
    &+ \frac{x^\dagger_\alpha y^\dagger_{N_i}}{t-m^2_\eta}(2p_1-p3)^\mu  \mathcal{C}_3 + \left(\frac{x^\dagger_\alpha \bar{\sigma}^\mu(p_1+p_2)\sigma y^\dagger_{N_i}}{s} + \frac{x^\dagger_\alpha y^\dagger_{N_i}(2p_1-p3)^\mu}{(t-m^2_\eta)}\right).
\end{align}
The cross term coming from the above two expressions will give us 
\begin{align}
    \Im[\mathcal{A}_0\mathcal{A}_1]|\mathcal{W}|^2 &= \left[ m_{N_i}m_{N_j}\frac{2m^4_{N_j} + 2m^2_\eta t + s(s+t) - m^2_{N_i}(2m^2_\eta + 3s + 2t)}{s(t-m^2_\eta)} \right. \nonumber \\
    &+ \left. 2m_{N_i}m_{N_j}\frac{(t-m^2_\eta)}{s}\right]\Im[\mathcal{C}_1] + \left[2m_{N_i}m_{N_j}\frac{m^2_\eta -s -t}{s} \right.\nonumber \\
    &+\left. m_{N_i}m_{N_j}\frac{2m^4_{N_i} + t(2m^2_\eta - s) - m^2_{N_i}(2m^2_\eta + s + 2t)}{s(m^2_\eta - t)}\right]\Im[\mathcal{C}_2] \nonumber \\
    &+ \left[2m_{N_i}\frac{(s+t-m^2_{N_i})(m^2_\eta - t)}{(t-m^2_\eta)^2} \right. \nonumber \\
    &+ \left. m_{N_i}\frac{(m^4_{N_i} + t(2m^2_\eta - s) - m^2_{N_i}(2m^2_\eta +s+2t))}{s(m^2_\eta - t)}\right]\Im[\mathcal{C}_3] \nonumber \\
    &+ \left[\frac{2(m^2_{N_i}-t)(m^2_\eta+t)}{(m^2_\eta-t)^2}-\frac{2(s+t-m^2_\eta)}{s} \right. \nonumber \\
    &+\left.\frac{2m^2_{N_i}+m_{N_i}(2m^2_\eta-s)t-m^3_{N_i}(2m^2_\eta + s + 2t)}{s(m^2_\eta - t)}\right]\frac{m_{N_i}m_{N_j}\Gamma_j}{16\pi^2((m^2_{N_i}-m^2_{N_j})^2 + m^2_j\Gamma^2_j)},
\end{align}
where
\begin{align}
    \mathcal{C}_1 &= \frac{(s-m^2_\eta-m^2_{N_i})}{\mathcal{D}}\left[\mathcal{B}_0(m^2_\eta,0,m^2_{N_j}) - \mathcal{B}_0(s,m^2_{N_j},m^2_\eta) \right. \nonumber \\
    &+ \left.(m^2_\eta - m^2_{N_i})C_0(m^2_\eta,m^2_{N_i},(m^2_\eta+m^2_{N_i}-s)/2,0,m^2_\eta,m^2_{N_j})\right], \nonumber \\
    &- \frac{2m^2_{N_i}}{\mathcal{D}}\left[\mathcal{B}_0(s,m^2_\eta,m^2_{N_i}) - \mathcal{B}_0(m^2_{N_i},0,m^2_\eta) \right. \nonumber \\
    &+ \left.(m^2_{N_i} - m^2_{N_j})C_0(m^2_\eta,m^2_{N_i},(m^2_\eta+m^2_{N_i}-s)/2,0,m^2_\eta,m^2_{N_j})\right] \nonumber 
\end{align}
\begin{align}
    \mathcal{C}_2 &= \frac{(s-m^2_\eta-m^2_{N_i})}{\mathcal{D}}\left[\mathcal{B}_0(s,m^2_\eta,m^2_{N_j}) - \mathcal{B}_0(m^2_{N_i},0,m^2_\eta)\right. \nonumber \\
    &+ \left.(m^2_{N_i} - m^2_{N_j})C_0(m^2_\eta,m^2_{N_i},(m^2_\eta+m^2_{N_i}-s)/2,0,m^2_\eta,m^2_{N_j})\right], \nonumber \\
    &- \frac{2m^2_\eta}{\mathcal{D}}\left[\mathcal{B}_0(m^2_\eta,0,m^2_{N_j}) - \mathcal{B}_0(s,m^2_\eta,m^2_{N_j}) \right. \nonumber \\
    &+ \left.(m^2_\eta - m^2_{N_i})C_0(m^2_\eta,m^2_{N_i},(m^2_\eta+m^2_{N_i}-s)/2,0,m^2_\eta,m^2_{N_j})\right], \nonumber 
\end{align}
\begin{align}
    \mathcal{C}_3 &= \frac{1}{m^2_{N_i}-t}\left[\mathcal{B}_0(t,0,m^2_{N_j}) - \mathcal{B}_0(m^2_{N_j},m^2_\eta,0) \right. \nonumber \\
    &+ \left.(m^2_\eta - m^2_{N_j} + t - m^2_{N_i})\mathcal{C}_0(m^2_\eta,m^2_{N_i},(m^2_{N_i}-t)/2,m^2_\eta,m^2_{N_j},0)\right], \nonumber \\
    \mathcal{D} &= (s-m^2_\eta - m^2_{N_i})^2 - 4m^2_\eta m^2_{N_i}.
\end{align}
where the scalar integral $\mathcal{B}_0$ and $\mathcal{C}_0$ is given by
\begin{align}
    \mathcal{B}_0(p^2,m^2_1,m^2_3) &= \mu^{2\epsilon}\int \frac{d^d}{(2\pi)^d} \frac{1}{(l^2-m^2_1)((l+p)^2-m^2_2)}, \nonumber \\
    \mathcal{C}_0(p^2_1,p^2_2,p1.p2,m^2_1,m^2_2,m^2_3) &= \mu^{2\epsilon}\int \frac{d^d}{(2\pi)^d}\frac{1}{(l^2-m^2_1)((l+p_1)^2-m^2_2)((l+p_2)^2-m^2_3)}.
\end{align}
Now the thermally averaged cross section is given as 
\begin{align}
    \langle \sigma_\delta v\rangle &= \frac{T}{32\pi m^2_\eta m^2_{N_i} K_2(m_\eta/T)K_2(m_{N_i}/T)}\int^\infty_{(m_\eta + m_{N_i})^2}\int^{t_+}_{t_-}ds dt |\delta|^2\frac{p_{in}p_{out}K_1(\sqrt{s}/T)}{\sqrt{s}}, \\
    p_{in} &= \frac{1}{2}\frac{\lambda^{1/2}(s,m^2_\eta,m^2_{N_i})}{\sqrt{s}}; \quad p_{out} = \frac{\sqrt{s}}{2}; \quad t_{\pm} = \frac{1}{2}(m^2_\eta + m^2_{N_i} - s \pm p_{in}p_{out}). \nonumber
\end{align}
Now taking the s-wave approximation of the above expression
\begin{align}
    \langle \sigma_\delta v \rangle &\simeq \sum_{ij}\frac{\Im[(yy^\dagger)^2_{ij}]}{8\pi(1+x_{N_i})}\widetilde{\epsilon}_{ij}(x^2_i(x_i-3)),\\
 \widetilde{\epsilon}_{ij}   &= \frac{\sqrt{x_j}}{6 x_i
   \left(-x_i^{3/2}+x_i (x_j-2)+\sqrt{x_i} x_j+1\right)^2(\sqrt{x_i}-3)} \left(x_i^{7/2} (3 x_j+1) +\sqrt{x_i} (3 x_j+5)+1 \right.\nonumber \\
   &- \left.3 x_i^{5/2} \left(x_j \left(D+(x_j-3) x_j+4\right)  -  3 D -2\right)-3 x_i^{3/2}
   \left(2 \left(D+3\right) + x_j \left(x_j \left(D+x_j+1\right)-D-4\right)\right) \right. \nonumber \\
   &- \left. x_i^4+f^3 \left(3
   D+3 x_j^2+11\right) - 3 x_i^2 \left(x_j \left(D+2 (x_j-1) x_j+2\right) - D +6\right)+x_i \left(1-3
   x_j \left(D+x_j-4\right)\right) \right) \nonumber \\
   &+ \frac{\sqrt{x_j}}{4 x_i}\left(\sqrt{x_i}-1+\frac{\sqrt{x_j}}{(1+\sqrt{x_i})^2}(\sqrt{x_i}-1+r_j)\left(\log\left(\frac{1+\sqrt{x_i}x_j}{x_i(1+\sqrt{x_i})}\right)-\log\left(\frac{1+x_i+x^{3/2}_i+\sqrt{x_i}x_j}{x_i(1+\sqrt{x_i})} \right)\right.\right. \nonumber \\
   &+ \left.\left. \log\left(1+\frac{1+\sqrt{x_i}}{\sqrt{x_i}(\sqrt{x_i}-1+x_i+x_j)}\right)\right)\right) + \frac{\sqrt{x_i} x_j\tilde{\Gamma}_j}{\pi((x_i-x_j)^2 + x_j\tilde{\Gamma}^2_j)}, \label{eq:coanneps} \\
   D &= \sqrt{(x_i-x_j) \left(x_i+4 \sqrt{x_i}-x_j+4\right)}, \qquad
  x_l = \frac{M^2_{N_l}}{m^2_\eta}, \qquad \tilde{\Gamma}_j = \frac{\Gamma_j}{m_\eta}. \nonumber
\end{align}

\acknowledgments
One of the authors, DB acknowledges the hospitality and facilities provided by School of Liberal Arts, Seoul-Tech, Korea where this work was completed. SK and AD were supported by the National Research Foundation of Korea (NRF) grants (2017K1A3A7A09016430, 2017R1A2B4006338).

\bibliographystyle{apsrev}

\begin{thebibliography}{10}

\bibitem{Aghanim:2018eyx}
{\bf Planck} Collaboration, N.~Aghanim et~al., {\it {Planck 2018 results. VI.
  Cosmological parameters}},  \href{http://arxiv.org/abs/1807.06209}{{\tt
  arXiv:1807.06209}}.

\bibitem{Weinberg:1979bt}
S.~Weinberg, {\it {Cosmological Production of Baryons}},  {\em Phys. Rev.
  Lett.} {\bf 42} (1979) 850--853.

\bibitem{Kolb:1979qa}
E.~W. Kolb and S.~Wolfram, {\it {Baryon Number Generation in the Early
  Universe}},  {\em Nucl. Phys.} {\bf B172} (1980) 224. [Erratum: Nucl.
  Phys.B195,542(1982)].

\bibitem{Fukugita:1986hr}
M.~Fukugita and T.~Yanagida, {\it {Baryogenesis Without Grand Unification}},
  {\em Phys. Lett.} {\bf B174} (1986) 45--47.


\bibitem{Patrignani:2016xqp}
{\bf Particle Data Group} Collaboration, C.~Patrignani et~al., {\it {Review of
  Particle Physics}},  {\em Chin. Phys.} {\bf C40} (2016), no.~10 100001.

\bibitem{Nussinov:1985xr}
S.~Nussinov, {\it {TECHNOCOSMOLOGY: COULD A TECHNIBARYON EXCESS PROVIDE A
  'NATURAL' MISSING MASS CANDIDATE?}},  {\em Phys. Lett.} {\bf 165B} (1985)
  55--58.

\bibitem{Davoudiasl:2012uw}
H.~Davoudiasl and R.~N. Mohapatra, {\it {On Relating the Genesis of Cosmic
  Baryons and Dark Matter}},  {\em New J. Phys.} {\bf 14} (2012) 095011,
  [\href{http://arxiv.org/abs/1203.1247}{{\tt arXiv:1203.1247}}].

\bibitem{Petraki:2013wwa}
K.~Petraki and R.~R. Volkas, {\it {Review of asymmetric dark matter}},  {\em
  Int. J. Mod. Phys.} {\bf A28} (2013) 1330028,
  [\href{http://arxiv.org/abs/1305.4939}{{\tt arXiv:1305.4939}}].

\bibitem{Zurek:2013wia}
K.~M. Zurek, {\it {Asymmetric Dark Matter: Theories, Signatures, and
  Constraints}},  {\em Phys. Rept.} {\bf 537} (2014) 91--121,
  [\href{http://arxiv.org/abs/1308.0338}{{\tt arXiv:1308.0338}}].

\bibitem{Yoshimura:1978ex}
M.~Yoshimura, {\it {Unified Gauge Theories and the Baryon Number of the
  Universe}},  {\em Phys. Rev. Lett.} {\bf 41} (1978) 281--284. [Erratum: Phys.
  Rev. Lett.42,746(1979)].

\bibitem{Barr:1979wb}
S.~M. Barr, {\it {Comments on Unitarity and the Possible Origins of the Baryon
  Asymmetry of the Universe}},  {\em Phys. Rev.} {\bf D19} (1979) 3803.

\bibitem{Baldes:2014gca}
I.~Baldes, N.~F. Bell, K.~Petraki, and R.~R. Volkas, {\it
  {Particle-antiparticle asymmetries from annihilations}},  {\em Phys. Rev.
  Lett.} {\bf 113} (2014), no.~18 181601,
  [\href{http://arxiv.org/abs/1407.4566}{{\tt arXiv:1407.4566}}].

\bibitem{Cui:2011ab}
Y.~Cui, L.~Randall, and B.~Shuve, {\it {A WIMPy Baryogenesis Miracle}},  {\em
  JHEP} {\bf 04} (2012) 075, [\href{http://arxiv.org/abs/1112.2704}{{\tt
  arXiv:1112.2704}}].

\bibitem{Bernal:2012gv}
N.~Bernal, F.-X. Josse-Michaux, and L.~Ubaldi, {\it {Phenomenology of WIMPy
  baryogenesis models}},  {\em JCAP} {\bf 1301} (2013) 034,
  [\href{http://arxiv.org/abs/1210.0094}{{\tt arXiv:1210.0094}}].

\bibitem{Bernal:2013bga}
N.~Bernal, S.~Colucci, F.-X. Josse-Michaux, J.~Racker, and L.~Ubaldi, {\it {On
  baryogenesis from dark matter annihilation}},  {\em JCAP} {\bf 1310} (2013)
  035, [\href{http://arxiv.org/abs/1307.6878}{{\tt arXiv:1307.6878}}].

\bibitem{Kumar:2013uca}
J.~Kumar and P.~Stengel, {\it {WIMPy Leptogenesis With Absorptive Final State
  Interactions}},  {\em Phys. Rev.} {\bf D89} (2014), no.~5 055016,
  [\href{http://arxiv.org/abs/1309.1145}{{\tt arXiv:1309.1145}}].

\bibitem{Racker:2014uga}
J.~Racker and N.~Rius, {\it {Helicitogenesis: WIMPy baryogenesis with sterile
  neutrinos and other realizations}},  {\em JHEP} {\bf 11} (2014) 163,
  [\href{http://arxiv.org/abs/1406.6105}{{\tt arXiv:1406.6105}}].

\bibitem{Dasgupta:2016odo}
A.~Dasgupta, C.~Hati, S.~Patra, and U.~Sarkar, {\it {A minimal model of TeV
  scale WIMPy leptogenesis}},  \href{http://arxiv.org/abs/1605.01292}{{\tt
  arXiv:1605.01292}}.

\bibitem{Ma:2006km}
E.~Ma, {\it {Verifiable radiative seesaw mechanism of neutrino mass and dark
  matter}},  {\em Phys. Rev.} {\bf D73} (2006) 077301,
  [\href{http://arxiv.org/abs/hep-ph/0601225}{{\tt hep-ph/0601225}}].


\bibitem{Racker:2013lua} 
  J.~Racker,
  JCAP {\bf 1403}, 025 (2014)
  doi:10.1088/1475-7516/2014/03/025
  [arXiv:1308.1840 [hep-ph]].

\bibitem{Hugle:2018qbw}
T.~Hugle, M.~Platscher, and K.~Schmitz, {\it {Low-Scale Leptogenesis in the
  Scotogenic Neutrino Mass Model}},
  \href{http://arxiv.org/abs/1804.09660}{{\tt arXiv:1804.09660}}.
  
  \bibitem{Borah:2018rca}
D. Borah, P. S. Bhupal Dev and A. Kumar, {\it {TeV scale leptogenesis, inflaton dark matter and neutrino mass in a scotogenic model}}, {\em Phys. Rev.} {\bf D99} (2019) 055012,
  \href{http://arxiv.org/abs/1810.03645}{{\tt arXiv:1810.03645}}.
  
  \bibitem{Casas:2001sr}
J.~A. Casas and A.~Ibarra, {\it {Oscillating neutrinos and muon $\rightarrow$ e,
  gamma}},  {\em Nucl. Phys.} {\bf B618} (2001) 171--204,
  [\href{http://arxiv.org/abs/hep-ph/0103065}{{\tt hep-ph/0103065}}].

\bibitem{Toma:2013zsa}
T.~Toma and A.~Vicente, {\it {Lepton Flavor Violation in the Scotogenic
  Model}},  {\em JHEP} {\bf 01} (2014) 160,
  [\href{http://arxiv.org/abs/1312.2840}{{\tt arXiv:1312.2840}}].
  
\bibitem{Salvio:2011sf}
  A.~Salvio, P.~Lodone and A.~Strumia,
  JHEP {\bf 1108} (2011) 116
  doi:10.1007/JHEP08(2011)116
  [arXiv:1106.2814 [hep-ph]].
  
\bibitem{Ibarra:2003up}
A. Ibarra and G. G. Ross, {\it Neutrino phenomenology: The case of two right-handed neutrinos}, {\em Phys. Lett.} {\bf B591} (2004) 285--296,  [\href{http://arxiv.org/abs/hep-ph/0312138}{{\tt hep-ph/0312138}}].

\bibitem{Mahanta:2019gfe}
D. Mahanta and D. Borah, {\it Fermion Dark Matter with $N_2$ Leptogenesis in Minimal Scotogenic Model},  [\href{http://arxiv.org/abs/1906.03577}{{\tt 1906.03577}}].


\bibitem{Barbieri:2006dq}
R.~Barbieri, L.~J. Hall, and V.~S. Rychkov, {\it {Improved naturalness with a
  heavy Higgs: An Alternative road to LHC physics}},  {\em Phys. Rev.} {\bf
  D74} (2006) 015007, [\href{http://arxiv.org/abs/hep-ph/0603188}{{\tt
  hep-ph/0603188}}].

\bibitem{LopezHonorez:2006gr}
L.~Lopez~Honorez, E.~Nezri, J.~F. Oliver, and M.~H.~G. Tytgat, {\it {The Inert
  Doublet Model: An Archetype for Dark Matter}},  {\em JCAP} {\bf 0702} (2007)
  028, [\href{http://arxiv.org/abs/hep-ph/0612275}{{\tt hep-ph/0612275}}].


\bibitem{Griest:1990kh}
K.~Griest and D.~Seckel, {\it {Three exceptions in the calculation of relic
  abundances}},  {\em Phys. Rev.} {\bf D43} (1991) 3191--3203.

\bibitem{Edsjo:1997bg}
J.~Edsjo and P.~Gondolo, {\it {Neutralino relic density including
  coannihilations}},  {\em Phys. Rev.} {\bf D56} (1997) 1879--1894,
  [\href{http://arxiv.org/abs/hep-ph/9704361}{{\tt hep-ph/9704361}}].


\bibitem{Esteban:2018azc} 
  I.~Esteban, M.~C.~Gonzalez-Garcia, A.~Hernandez-Cabezudo, M.~Maltoni and T.~Schwetz,
  arXiv:1811.05487 [hep-ph].

\bibitem{Staub:2013tta}
F.~Staub, {\it {SARAH 4 : A tool for (not only SUSY) model builders}},  {\em
  Comput. Phys. Commun.} {\bf 185} (2014) 1773--1790,
  [\href{http://arxiv.org/abs/1309.7223}{{\tt arXiv:1309.7223}}].

\bibitem{Barducci:2016pcb}
D.~Barducci, G.~Belanger, J.~Bernon, F.~Boudjema, J.~Da~Silva, S.~Kraml,
  U.~Laa, and A.~Pukhov, {\it {Collider limits on new physics within
  micrOMEGAs 4.3}},  {\em Comput. Phys. Commun.} {\bf 222} (2018) 327--338,
  [\href{http://arxiv.org/abs/1606.03834}{{\tt arXiv:1606.03834}}].

\bibitem{sph1}
 V.~A.~Kuzmin, V.~A.~Rubakov and M.~E.~Shaposhnikov,
  Phys.\ Lett.\  {\bf 155B}, 36 (1985).

\bibitem{sph2}
  M.~D'Onofrio, K.~Rummukainen and A.~Tranberg,
  Phys.\ Rev.\ Lett.\  {\bf 113}, no. 14, 141602 (2014)
  doi:10.1103/PhysRevLett.113.141602
  [arXiv:1404.3565 [hep-ph]].

\bibitem{Porod:2011nf}
W.~Porod and F.~Staub, {\it {SPheno 3.1: Extensions including flavour,
  CP-phases and models beyond the MSSM}},  {\em Comput. Phys. Commun.} {\bf
  183} (2012) 2458--2469, [\href{http://arxiv.org/abs/1104.1573}{{\tt
  arXiv:1104.1573}}].

\bibitem{Cui:2009xq} 
  Y.~Cui, D.~E.~Morrissey, D.~Poland and L.~Randall,
  JHEP {\bf 0905}, 076 (2009)
  doi:10.1088/1126-6708/2009/05/076
  [arXiv:0901.0557 [hep-ph]].

\bibitem{Arhrib:2015dez}
A.~Arhrib, C.~Boehm, E.~Ma, and T.-C. Yuan, {\it {Radiative Model of Neutrino
  Mass with Neutrino Interacting MeV Dark Matter}},  {\em JCAP} {\bf 1604}
  (2016), no.~04 049, [\href{http://arxiv.org/abs/1512.08796}{{\tt
  arXiv:1512.08796}}].

\bibitem{thooft}
G.~'t Hooft, Naturalness, chiral symmetry and spontaneous chiral symmetry breaking, NATO Sci. Ser, B, 59, (1980) 135.

\bibitem{Aprile:2018dbl}
E.~Aprile et~al., {\it {Dark Matter Search Results from a One Tonne$\times$Year
  Exposure of XENON1T}},  \href{http://arxiv.org/abs/1805.12562}{{\tt
  arXiv:1805.12562}}.

\bibitem{Ahnen:2016qkx}
{\bf Fermi-LAT, MAGIC} Collaboration, M.~L. Ahnen et~al., {\it {Limits to dark
  matter annihilation cross-section from a combined analysis of MAGIC and
  Fermi-LAT observations of dwarf satellite galaxies}},  {\em JCAP} {\bf 1602}
  (2016), no.~02 039, [\href{http://arxiv.org/abs/1601.06590}{{\tt
  arXiv:1601.06590}}].

\bibitem{TheMEG:2016wtm}
{\bf MEG} Collaboration, A.~M. Baldini et~al., {\it {Search for the lepton
  flavour violating decay $\mu ^+ \rightarrow \mathrm {e}^+ \gamma $ with the
  full dataset of the MEG experiment}},  {\em Eur. Phys. J.} {\bf C76} (2016),
  no.~8 434, [\href{http://arxiv.org/abs/1605.05081}{{\tt arXiv:1605.05081}}].

\bibitem{Bellgardt:1987du}
{\bf SINDRUM} Collaboration, U.~Bellgardt et~al., {\it {Search for the Decay
  $\mu^+ \rightarrow e^+ e^+ e^-$ }},  {\em Nucl. Phys.} {\bf B299} (1988) 1--6.

\bibitem{Dohmen:1993mp}
{\bf SINDRUM II} Collaboration, C.~Dohmen et~al., {\it {Test of lepton flavor
  conservation in $\mu \rightarrow e$ conversion on titanium}},  {\em Phys. Lett.} {\bf
  B317} (1993) 631--636.

\bibitem{Cannoni:2015wba} 
  M.~Cannoni,
  Eur.\ Phys.\ J.\ C {\bf 76}, no. 3, 137 (2016)
  doi:10.1140/epjc/s10052-016-3991-2
  [arXiv:1506.07475 [hep-ph]].

\end{thebibliography}
\providecommand{\href}[2]{#2}\begingroup\raggedright
\endgroup

\end{document}